\documentclass[floatfix,aps,reprint,prb]{revtex4-1}
\usepackage{graphicx} \usepackage{amsmath, amssymb, bm}
\usepackage{dcolumn} \usepackage{color}
\usepackage{fix-cm}
\usepackage{times}
\usepackage{float}

\usepackage{multirow}
\usepackage{sidecap}
\usepackage{diagbox}
\usepackage{marginnote}
\usepackage{ulem} %package for the strike out \sout and \uwave
\usepackage{rotating}

\newcommand{\rzcm}{cm$^{-1}$}
\renewcommand{\emph}[1]{\textit{#1}}

\begin{document}
\title{Infrared dielectric functions and Brillouin zone center phonons of $\alpha$-Ga$_2$O$_3$ compared to $\alpha$-Al$_2$O$_3$}
\author{Megan Stokey}
\email{mstokey@huskers.unl.edu}
\homepage{http://ellipsometry.unl.edu}
\affiliation{Department of Electrical and Computer Engineering, University of Nebraska-Lincoln, Lincoln, NE 68588, USA}
\author{Rafa\l{} Korlacki}
\affiliation{Department of Electrical and Computer Engineering, University of Nebraska-Lincoln, Lincoln, NE 68588, USA}
\author{Sean Knight}
\affiliation{Terahertz Materials Analysis Center and Center for III-N Technology, C3NiT -- Janz\`{e}n, Department of Physics, Chemistry and Biology (IFM), Link\"{o}ping University, 58183 Link\"{o}ping, Sweden}
\author{Steffen Richter}
\affiliation{Terahertz Materials Analysis Center and Center for III-N Technology, C3NiT -- Janz\`{e}n, Department of Physics, Chemistry and Biology (IFM), Link\"{o}ping University, 58183 Link\"{o}ping, Sweden}
\author{Riena Jinno}
\affiliation{School of Electrical and Computer Engineering, Cornell University, Ithaca, NY 14853, USA}
\affiliation{Department of Electronic Science and Engineering, Kyoto University, Kyoto, 615-8510, Japan}
\author{Yongjin Cho}
\affiliation{School of Electrical and Computer Engineering, Cornell University, Ithaca, NY 14853, USA}
\author{Huili Grace Xing}
\affiliation{School of Electrical and Computer Engineering, Cornell University, Ithaca, NY 14853, USA}
\affiliation{Department of Material Science and Engineering, Cornell University, Ithaca, NY 14853, USA}
\author{Debdeep Jena}
\affiliation{School of Electrical and Computer Engineering, Cornell University, Ithaca, NY 14853, USA}
\affiliation{Department of Material Science and Engineering, Cornell University, Ithaca, NY 14853, USA}
\author{Matthew Hilfiker}
\affiliation{Department of Electrical and Computer Engineering, University of Nebraska-Lincoln, Lincoln, NE 68588, USA}
\author{Vanya Darakchieva}
\affiliation{Terahertz Materials Analysis Center and Center for III-N Technology, C3NiT -- Janz\`{e}n, Department of Physics, Chemistry and Biology (IFM), Link\"{o}ping University, 58183 Link\"{o}ping, Sweden}
\author{Yuichi Oshima}
\affiliation{Optical Single Crystals Group, National Institute for Materials Science, Tsukuba, Ibaraki 3050044, Japan}
\author{Kamruzzaman Khan}
\affiliation{Electrical Engineering and Computer Science Department, University of Michigan, Ann Arbor, MI 48109, USA }
\author{Elaheh Ahmadi}
\affiliation{Electrical Engineering and Computer Science Department, University of Michigan, Ann Arbor, MI 48109, USA }
\author{Mathias Schubert}
\affiliation{Department of Electrical and Computer Engineering, University of Nebraska-Lincoln, Lincoln, NE 68588, USA}
\affiliation{Terahertz Materials Analysis Center and Center for III-N Technology, C3NiT -- Janz\`{e}n, Department of Physics, Chemistry and Biology (IFM), Link\"{o}ping University, 58183 Link\"{o}ping, Sweden}

\date{\today}

\begin{abstract}
 We determine the anisotropic dielectric functions of rhombohedral $\alpha$-Ga$_2$O$_3$ by far-infrared and infrared generalized spectroscopic ellipsometry and derive all transverse optical and longitudinal optical phonon mode frequencies and broadening parameters. We also determine the high frequency and static dielectric constants. We perform density functional theory computations and determine the phonon dispersion for all branches in the Brillouin zone, and we derive all phonon mode parameters at the Brillouin zone center including Raman-active, infrared-active, and silent modes. Excellent agreement is obtained between our experimental and computation results as well as among all previously reported partial information from experiment and theory. We also compute the same information for $\alpha$-Al$_2$O$_3$, the binary parent compound for the emerging alloy of $\alpha$-(Al$_{x}$Ga$_{1-x}$)$_2$O$_3$, and use results from previous investigations [Schubert, Tiwald, and Herzinger, Phys. Rev. B 61, 8187 (2000)] to compare all properties among the two isostructural compounds. From both experimental and theoretical investigations we compute the frequency shifts of all modes between the two compounds. Additionally, we calculate overlap parameters between phonon mode eigenvectors and discuss the possible evolution of all phonon modes into the ternary alloy system and whether modes may form single mode or more complex mode behaviors. 
\end{abstract}

\maketitle
\section{Introduction}
\begin{figure}%[hbt]
\centering
\includegraphics[width=.75\linewidth]{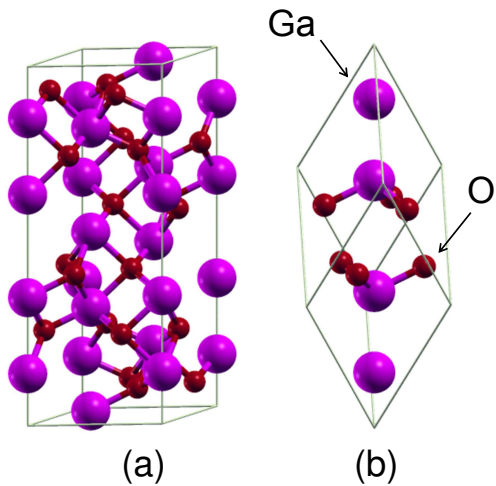}
\caption{\label{fig:cells} Crystallographic unit cells for $\alpha$-Ga$_2$O$_3$: (a) hexagonal; (b) rhombohedral. }
\end{figure}

With the increasing demand to improve the efficiency of electric power conversion using semiconductor devices, the monoclinic ($\beta$) phase of Ga$_2$O$_3$ has emerged as a new promising material. Due to its ultra-wide bandgap energy of about 5~eV, the resulting large breakdown electric field surpasses SiC and GaN by a factor of two.\cite{HigashiwakiAPL2018Guest,Mock_2017Ga2O3,PeartonAPRev2018bGO} It is known that at least five polymorphs of Ga$_2$O$_3$ exist: $\alpha$, $\beta$, $\gamma$, $\delta$, and $\epsilon$ phases.\cite{RoyJACS1952} Only the $\beta$ phase is stable and can be grown conveniently from the melt.\cite{ChaseJACS} Epitaxial growth recently succeeded in stabilizing the $\alpha$ phase, $\alpha$-Ga$_2$O$_3$ (aGO), as thin films in heteroepitaxial growth on sapphire ($\alpha$-Al$_2$O$_3$; aAO) substrates at low-temperatures.\cite{ShinoharaJJAP,ShinoharaJJAP,FUJITA2014,Kawaharamura_2012} The impact of dry etching conditions necessary for device fabrication was investigated.\cite{Jian_2019} Schottky barrier diode\cite{Oda_2016} and Sn-doped MESFET\cite{Dang_2015} operations were demonstrated. Now heteroepitaxial growth raises many possibilities for device design because the substrates, such as sapphire, are commercially available. The structure of the $\alpha$ phase (rhombohedral, $R \bar{3}c$, space group 167, Figure~\ref{fig:cells}) is identical with the structure of sapphire (aAO, also referred to as corrundum) and possesses higher symmetry than the thermodynamically stable $\beta$ phase. The higher symmetry of aGO is seen favorable over monoclinic gallium oxide because material properties and device design will simplify. For example, $\beta$-Ga$_2$O$_3$ possesses only one high symmetry direction and lattice vibrations within the monoclinic plane do not align with a low-indexed crystallographic direction.\cite{SchubertbGOphononchapter,PhysRevB.93.125209} Phonon-plasmon scattering in $\beta$-Ga$_2$O$_3$ strongly affects the electrical transport, where coupled longitudinal (LO) phonon modes depend in their scattering directions on the Fermi energy level, and acoustic and optical mode anisotropy unfavorably limits electron mobility.\cite{SchubertAPLLPPbGO2019,ghosh_singisetti_2017,KumarAPL2020,PoncePRR2020} Likewise, the band-to-band transitions consist of three transitions whose polarization directions do not all coincide with high packing density lattice directions in the monoclinic plane.\cite{Mock_2017Ga2O3,Sturm_2016} The three fundamental transition energies differ by as much as 600~meV ($E^{ac}_0=5.041$~eV, $E^{ac}_1=5.401$~eV, $E^{b}_0=5.641$~eV)\cite{Mock_2017Ga2O3} which lead to strongly differing direction dependent band offsets, e.g., with $\beta$-(Al,Ga)$_2$O$_3$ (bAGO).\cite{MuAPL2021} Likewise, the static dielectric constant is strongly direction dependent varying from 10.19 along crystallographic axis $\mathbf{a}$ to 10.6 along axis $\mathbf{b}$, and to 12.4 along axis $\mathbf{c}$.\cite{GopalanAPL2020} This variation complicates device design. The effective electron mass of $\beta$-Ga$_2$O$_3$ has only small anisotropy and was determined to be 0.267$m_e$.\cite{KnightbGOOHE2018} On the contrary, aGO is optically uniaxial and is characterized by only two fundamental band-to-band transitions polarized along high symmetry directions.\cite{Hilfiker_21aGO} Furthermore, aGO also possesses only half the number of phonon modes in the set of Brillouin zone center lattice vibrations whose displacement directions all align with high symmetry lattice directions.

While aGO and aAO share the same crystal structure, alloys of $\alpha$-(Al,Ga)$_2$O$_3$ (aAGO) constitute an emerging ultra-wide bandgap semiconductor system with tunable large bandgap energies ($E_g$ $\approx$5.4~eV – 8.8~eV) and high critical electric field strengths. Such properties render aGO and aAGO extraordinarily attractive for next generation power, radio frequency, and deep-UV optoelectronic devices.\cite{HigashiwakiAPL2018Guest} As another advantage, aGO and aAGO alloys could potentially be integrated with corundum structured $p$-type oxides such as Rh$_2$O$_3$ or Ir$_2$O$_3$. Such combination may help to overcome the typical challenge of unipolar operation in ultra-wide bandgap semiconductors. The extremely wide bandgap may also offer new possibilities of using controlled defect centers for potential quantum computing applications.\cite{SonAPL2020}     

\begin{table}\centering
\caption{Previously reported relevant properties for $\alpha$-Ga$_2$O$_3$ and $\alpha$-Al$_2$O$_3$.}
\label{tab:introduction}
\begin{ruledtabular}
\begin{tabular}{{c}{c}{c}{c}}
Parameter & Method  & Ref. aGO & Ref. aAO\\
\hline
$\varepsilon_{\infty}$ & Calc.  & \onlinecite{HePRBaGOtheory2006, Furthmuller_2016,Hilfiker_21aGO, Feneberg19aGO,Sharma2021} &\onlinecite{Heid2000,Montanari2006,Griffin2018,mp1143}\\
& Exp. & \onlinecite{Hilfiker_21aGO,Hilfiker_21aAGO, Feneberg19aGO} & \onlinecite{SchubertPRB61_2000}\\
$\varepsilon_{\mathrm{DC}}$ & Calc. & \onlinecite{Sharma2021} &\onlinecite{Montanari2006,Griffin2018,mp1143}\\
& Exp. & Not avail.&\onlinecite{SchubertPRB61_2000} \\
Raman Phonon Modes & Calc. & \onlinecite{Cusco15,Sharma2021}&\onlinecite{Heid2000,Lodziana2003,Montanari2006,Griffin2018,mp1143}\\
& Exp. &\onlinecite{Cusco15, Feneberg18aGO} &\onlinecite{MunissoRamanExp,Pezzotti15}\\
TO Infrared Phonon Modes & Calc. & \onlinecite{Sharma2021, Feneberg18aGO}&\onlinecite{Heid2000,Lodziana2003,Montanari2006,Griffin2018,mp1143}\\
 & Exp. & \onlinecite{Feneberg19aGO}$^{a}$ &\onlinecite{SchubertPRB61_2000}\\
LO Infrared Phonon Modes & Calc. &\onlinecite{Sharma2021} & \onlinecite{Heid2000,Lodziana2003,Montanari2006,Griffin2018,mp1143}\\
& Exp. & Not avail. &\onlinecite{SchubertPRB61_2000} \\

\end{tabular}
\end{ruledtabular}
\begin{flushleft}
\footnotesize{$^a${Incomplete experimental set of TO
modes.}}\\
\end{flushleft}
\end{table}
The current state of knowledge about fundamental properties of aGO is still limited because high quality single crystalline material became available only recently through progress in heteroepitaxy. Most recently, Hilfiker~\textit{et al.} performed a wide spectral range analysis determining the anisotropic dielectric functions and fundamental band-to-band transitions for polarization directions parallel ($||$, extraordinary direction) and perpendicular ($\perp$, ordinary direction) to the lattice $\mathbf{c}$ axis in aGO using generalized spectroscopic ellipsometry from the near infrared (IR) to the vacuum ultraviolet (0.73~eV to 8.75~eV).\cite{Hilfiker_21aGO} The  band-to-band transitions for polarization perpendicular to the $\mathbf{c}$ axis, $E_{0,\perp}=5.46(6)$~eV and $E_{0,\perp}=6.041$~eV belong to $M_0$-type van Hove singularities, whereas a $M_1$-type van Hove singularity occurs for the band-to-band transition parallel to the $c$ axis ($E_{0,||}=5.442$~eV).\cite{Hilfiker_21aGO} Hybrid-level density functional theory was also used to elucidate the structure of both the valence and conduction bands and $\Gamma$-point effective mass parameters were provided along high symmetry directions. As an unusual property among conventional semiconductors, such as GaAs or GaN, the highly complex and anisotropic band structure of aGO results in excitonic contributions to the anisotropic band-to-band transitions. A small binding energy of 7~meV was found for the lowest ordinary transition and a hyperbolic (quasi 2-dimensional) exciton with large binding energy of 178~meV was observed for the $M_1$-type (extraordinary) critical point transition.\cite{Hilfiker_21aGO} In earlier work, Segura~\textit{et al.},\cite{SeguraPRMataGOEg2017} Kracht~\textit{et al.},\cite{KrachtPRApplaGO2018} and Feneberg~\textit{et al.}\cite{Feneberg18aGO} reported on more limited spectroscopic investigations and obtained partial bandgap energy information in agreement with Hilfiker~\textit{et al.} The anisotropic dielectric constant for photon energies far below the bandgap but far above optical phonon mode frequencies were obtained from IR ellipsometry measurements ($\varepsilon_{\infty,\perp}$=3.75 and $\varepsilon_{\infty,\parallel}$=3.64),\cite{Feneberg19aGO} which agreed reasonably well by extrapolation from the near IR ellipsometry analysis ($\varepsilon_{\infty,\bot}$ = 3.86, $\varepsilon_{\infty,\parallel}$ = 3.76) and direct Cauchy model analysis ($\varepsilon_{\infty,\bot}$ = 3.864, $\varepsilon_{\infty,\parallel}$ = 3.756) performed by Hilfiker~\textit{et al}.\cite{Hilfiker_21aGO,Hilfiker_21aAGO} 

Of fundamental importance for device design and operation is precise knowledge of phonon mode properties and static dielectric constants. A complete picture for these properties has not emerged yet. Table~\ref{tab:introduction} summarizes the present state of knowledge for $\alpha$-Ga$_2$O$_3$ while also presenting select references for comparable $\alpha$-Al$_2$O$_3$ parameters. While Raman investigations have lead to sufficient information for all Raman-active modes, only limited information is available for the IR-active modes. No experimental investigation of the static dielectric constant is available yet. Theoretical computations predict phonon and static permittivity. He~\textit{et al.} report calculated dielectric function spectra from 0 to 50~eV for both $\alpha$ and $\beta$ phases without phonon contributions,\cite{HePRBaGOtheory2006} which compare reasonably well with experiments that became available later.\cite{Mock_2017Ga2O3,Hilfiker_21aGO,Feneberg19aGO} Because phonon contributions were ignored, the isotropic dielectric constant for $\alpha$-Ga$_2$O$_3$ reported by He~\textit{et al.} is equivalent to the high frequency constant but was found slightly too low (3.03).\cite{HePRBaGOtheory2006} Solving the Bethe-Salpeter equation, Furthmueller and Bechstedt calculated the excitonic optical spectra of the two polymorphs and also obtained the high frequency dielectric constants (3.8).\cite{Furthmuller_2016} Cusc\'o~\textit{et al.} reported \textit{ab-initio} calculated and experimental frequencies for all symmetry-permitted Raman modes.\cite{Cusco15} Feneberg~\textit{et al.} also reported Raman experiments.\cite{Feneberg19aGO} Recently, Sharma and Singisetti predicted low-field electron transport in $\alpha$-Ga$_2$O$_3$ from density perturbation function theory calculations. This work also reported calculated phonon band structure, all zone center Raman-active, IR-active, and silent modes, including LO modes.\cite{Sharma2021} Values for high frequency dielectric constants ($\varepsilon_{\infty, \parallel}=4.46$, $\varepsilon_{\infty, \perp}=4.62$) and their anisotropy were provided. These values were somewhat larger than those found earlier from experiment. Feneberg~\textit{et al.} published data calculated by density functional theory (DFT) for all transverse optical (TO) phonon modes (4 modes with $E_{\mathrm{u}}$ and 2 modes with $A_{2\mathrm{u}}$  symmetry).\cite{Feneberg18aGO} Experimental investigations were reported by Feneberg~\textit{et al.} using spectroscopic ellipsometry in the IR range. All of the TO modes were identified except for the lowest $E_\mathrm{u}$ mode due to lack of access to the far-IR range.\cite{Feneberg18aGO,Feneberg19aGO} Akaiwa~\textit{et al.} currently report the highest mobility value achieved for tin-doped alpha phase gallium oxide at 65~cm$^{2}$/(Vs).\cite{Akaiwa20} Sharma and Singisetti predicted a low field isotropic average electron mobility of $\approx$~220~cm$^2$/(Vs) and suggested polar optical phonon scattering as the dominant limitation for electron density of $1.0 \times 10^{15}$~cm$^{-3}$.\cite{Sharma2021} Experimental data for the LO phonon modes have not been reported. The knowledge of these modes is of particular interest because with the accurate knowledge of the high frequency dielectric constants, the static dielectric constants can then be calculated using the Lyddane-Sachs-Teller (LST) relationship.\cite{Lyddane41} Sharma and Singisetti used their calculated phonon modes and predicted $\varepsilon_{\mathrm{DC}, \mathrm{||}}=18.67$ and $\varepsilon_{\mathrm{DC}, \perp}=12.98$.\cite{Sharma2021} An experimental determination of the static constants, e.g., using radio frequency capacitance measurements is unavailable presently. 

Very limited information is available currently for the alloy system aAGO. Jinno~\textit{et al.} determined the shift of the fundamental absorption edge from aGO to aAO from a set of epitaxial thin films.\cite{jinno2020crystal} Hilfiker~\textit{et al.} determined the high frequency dielectric constants and anisotropic indices of refraction across the alloy range.\cite{Hilfiker_21aAGO} No Raman or IR analysis of phonon modes has been reported. The influence of epitaxial strain, well known from the structurally similar AlGaN system is also widely unknown.\cite{PhysRevB.71.115329,PhysRevB.75.195217,PhysRevB.70.045411} The strain-stress relationships for all phonon modes were reported by Korlacki~\textit{et al.} recently for bGO,\cite{KorlackiPRB20} while no such information is available for aGO. Phonon mode properties evolve in a complex fashion from the binary parent compounds into the alloy system. These properties can be used to accurately assess strain, composition, and free charge carrier information, for example, which has been long investigated in the AlGaN system.\cite{Schoche2017} For aAGO, no such information is available yet.

In this work, we report a combined computational (DFT) and experimental (far-IR and IR generalized spectroscopic ellipsometry) investigation of the Brillouin zone center modes in $\alpha$-Ga$_2$O$_3$, in comparison with $\alpha$-Al$_2$O$_3$. A set of high structural quality single crystalline undoped aGO thin films is investigated for its anisotropic dielectric functions. From the dielectric function and DFT calculation results we identify and determine all TO and LO phonon modes with both $A_{2\mathrm{u}}$ and $E_{\mathrm{u}}$ symmetry. We further report all Raman and silent modes from our DFT calculations and compare all results with literature data where available. We apply the LST relationship and report the anisotropic static dielectric constants for aGO. We then compare the character of phonon modes in aGO with those in the isostructural compound aAO by calculation of the overlap parameters between all phonon mode eigenvectors. We find similarities and also point out differences. From our similarity analysis, we hypothesize about the mode character within the alloy system. These comparisons will become important when investigating optical phonons in the alloy system $\alpha$-(Al,Ga)$_2$O$_3$ in the future will impact future research and design for aAGO-based high power electronic devices.

\section{Methods}

\subsection{Structure}\label{structure}

$\alpha$-Ga$_2$O$_3$ crystallizes in the corundum structure ($R \bar{3}c$, space group 167), which is optically uniaxial and belongs to the trigonal crystal family. $\alpha$-Al$_2$O$_3$ (sapphire) has the same structure as $\alpha$-Ga$_2$O$_3$, and its terminology is often used synonymous to corundum. Figure~\ref{fig:cells} shows both, the hexagonal and the primitive rhombohedral unit cells of $\alpha$-Ga$_2$O$_3$. \footnote{All Figures with unit cells and mode displacement patters (the latter normalized to an arbitrary magnitude in order to visualize together modes with differing oscillator strengths), were prepared using XCrysDen~\cite{[][{. Code available from http://www.xcrysden.org.}]Kokalj1999} running under Silicon Graphics Irix 6.5.} The rhombohedral unit cell contains 10 atoms resulting in 30 phonon modes. In the Brillouin zone center 3 modes are acoustic (A$_{\mathrm{2u}}+2$E$_{\mathrm{u}}$) and the 27 optical modes are classified according to the irreducible representation:
\begin{equation*}
    \Gamma_{\mathrm{opt}} = 2A_{\mathrm{1g}} + 3A_{\mathrm{2g}} + 5E_{\mathrm{g}} + 2A_{\mathrm{1u}} + 2A_{\mathrm{2u}} + 4E_{\mathrm{u}},
\end{equation*}
with $E_{\mathrm{g}}$ and $E_{\mathrm{u}}$ modes being double degenerate. $A_{\mathrm{2u}}$ and $E_{\mathrm{u}}$ modes are IR-active, $A_{\mathrm{1g}}$ and $E_{\mathrm{g}}$ modes are Raman-active, and $A_{\mathrm{1u}}$ and $A_{\mathrm{2g}}$ modes are silent. Additionally, IR-active modes split into TO and LO modes forming associated pairs of TO-LO modes, where the LO modes are always larger than their associated TO modes, also known as the TO-LO rule.\cite{VFSbook,SchubertIRSEBook_2004}

\subsection{Density Functional Theory}

DFT calculations were performed using the open-source plane-wave code Quantum ESPRESSO (QE).\cite{[{Quantum ESPRESSO is available from http://www.qu\-an\-tum-es\-pres\-so.org. See also: }]GiannozziJPCM2009QE} We performed the calculations at two different levels of theory: the local density approximation (LDA) and the generalized-gradient-approximation (GGA). In general, it is expected that the frequencies of the phonon modes computed at the LDA level are closer to the experimental values than those computed at the GGA level. More subtle properties, such as Born effective charges and oscillator strengths, should be more accurate at the GGA level. For the LDA calculations we used the density functional of Perdew and Zunger (PZ)\cite{PerdewPRB1981}, the Bachelet-Hamann-Schl{\"u}ter (BHS)\cite{BHSPRB82} pseudpopotential for gallium, and Troullier-Martins\cite{TroullierPRB1991} pseudopotential for oxygen. For the GGA calculations we used the density functional of Perdew, Burke, and Ernzerhof (PBE)\cite{PBE1996} and norm-conserving Troullier-Martins pseudopotentials originally generated using the FHI98PP\cite{FuchsCPC1999,TroullierPRB1991} code and available in the Quantum ESPRESSO pseudopotentials library. Both pseudopotentials for gallium do not include semi-core 3$d$ states in the valence configuration. The same methods (specific density functionals as well as pseudopotentials) we have used before for studying phonon properties of $\beta$-Ga$_2$O$_3$, LDA in Ref.~\onlinecite{PhysRevB.93.125209} and GGA in, for example, Refs.~\onlinecite{SchubertPRB2019,KorlackiPRB20}. The details of the calculations and the specific parameters for obtaining the equilibrium cell of $\alpha$-Ga$_2$O$_3$ are the same as described in our previous paper on $\alpha$-Ga$_2$O$_3$.\cite{Hilfiker_21aGO} For reference, because $\alpha$-Ga$_2$O$_3$ belongs to the same space group as $\alpha$-Al$_2$O$_3$ (sapphire) we performed DFT calculations of $\alpha$-Al$_2$O$_3$ as well, at the GGA level only, with the exact same density functional, family of pseudopotentials, and convergence criteria as for $\alpha$-Ga$_2$O$_3$. The equilibrium cells for both materials, and all methods employed here, were used for subsequent phonon calculations. The phonon frequencies, Born effective charges, and IR intensities were computed at the $\Gamma$-point of the Brillouin zone using density functional perturbation theory,~\cite{BaroniRMP2001DFTPhonons} as implemented in the Quantum ESPRESSO package, with the convergence threshold for self-consistency of $1\times10^{-18}$ Ry. The parameters of the IR-active TO modes, as well as Raman-active and silent modes, were obtained from the dynamical matrices computed at the $\Gamma$-point. The parameters of the IR-active LO modes were obtained by setting a small displacement from the $\Gamma$-point in order to include the long-range Coulomb interactions of Born effective charges in the dynamical matrix (the so called non-analytical terms). In the case of $A_{\mathrm{2u}}$ modes, having their transition dipoles oriented along the main symmetry axis (the hexagonal $\mathbf{c}$ vector), the displacement for non-analytical terms was set in the direction of the symmetry axis. In the case of $E_{\mathrm{u}}$ modes, having their transition dipoles oriented normal to the main symmetry axis, the displacement for non-analytical terms can in principle be set in an arbitrary direction perpendicular to the main symmetry axis. Each $E_{\mathrm{u}}$ mode is double-degenerate, i.e., consists of two orthogonal modes, with mutually perpendicular transition dipoles, but are identical otherwise. Therefore, for the purpose of rendering the complete set of mode eigenvectors, we set the displacements for non-analytical terms in the directions of the respective TO modes. Each of such TO-LO mode pairs with $E_{\mathrm{u}}$ symmetry is identified in the Figures by indices $a$ and $b$, which should not be confused with unit cell parameters. In a similar manner we distinguish degenerate, Raman-active modes $E_{\mathrm{g}}$.

In addition to the phonon modes at the Brillouin zone center, we calculate the phonon dispersion along a high-symmetry path in the Brillouin zone. Dynamical matrices were calculated on a regular $4 \times 4 \times4$ grid in the Brillouin zone, and later used to produce real-space interatomic force constants, which in turn were used to plot the phonon dispersion along a path.

\subsection{Samples}

Two $\alpha$-Ga$_{2}$O$_3$ samples were investigated here. The first was heteroepitaxially grown as a (10$\bar{1}$0) thin-film on $m$-plane oriented $\alpha$-Al$_{2}$O$_3$ substrates via plasma-enhanced molecular beam epitaxy (PAMBE).\cite{jinno2020crystal} The substrate temperature was maintained at $650^\circ$C for growth. Prior to growth, the substrates were treated with an oxygen plasma for ten minutes at $T = 800^\circ$C. To create active oxygen species, a radio frequency plasma source was used with a oxygen flow rate of 0.5~sccm. The chamber pressure was kept at ~10$^{-5}$ Torr during the deposition. After growth was complete, x-ray reflectivity measurements were done to determine the epitaxial layer's thickness to be 51.8~nm. Atomic force microscopy indicated a root mean square roughness of 0.96~nm. By using asymmetrical reciprocal space map analysis, the $\alpha$-Ga$_{2}$O$_3$ films were determined to be completely strain-free and followed the $m$-plane orientation of the substrate. Note, $\alpha$-Ga$_{2}$O$_3$ and  $\alpha$-Al$_{2}$O$_3$ have different lattice constants and interfacial nucleation phase formation result in strain-free epitaxial material. Further information regarding growth and characterization can be found in Jinno~\emph{et al.}\cite{jinno2020crystal} 

The second sample investigated was an $\alpha$-Ga$_{2}$O$_3$ (0001) film grown on a $c$-plane oriented $\alpha$-Al$_{2}$O$_3$ substrate. This sample was grown by hydride vapor phase epitaxy (HVPE) at $520^{\circ}$C using GaCl and O$_2$ as the precursers in an atmospheric quartz reactor. Partial pressures of GaCl and O$_2$ were 0.125 and 1.25~kPa, respectively. The growth rate was approximately 12~$\mu$m/h, and the film thickness estimate from growth time (15 min) was approximately 3~$\mu$m. More information on similarly grown samples can be found in Oshima~\emph{et al.}\cite{Oshima_2020}

\subsection{Generalized Spectroscopic Ellipsometry}

Generalized spectroscopic ellipsometry (GSE) is a contactless measurement technique that can determine the anisotropic optical properties of a material. For each wavenumber, the Mueller matrix of the material is measured. The Mueller matrix relates the Stokes vector components before and after interaction with a sample, 
\begin{equation}
\left( {{\begin{array}{*{20}c}
 {S_{0} } \hfill \\ {S_{1} } \hfill \\  {S_{2} } \hfill \\  {S_{3} } \hfill \\
\end{array} }} \right)_{\mathrm{output}} =
\left( {{\begin{array}{*{20}c}
 {M_{11} } \hfill & {M_{12} } \hfill \ {M_{13} } \hfill & {M_{14} } \hfill \\
 {M_{21} } \hfill & {M_{22} } \hfill \ {M_{23} } \hfill & {M_{24} } \hfill \\
 {M_{31} } \hfill & {M_{32} } \hfill \ {M_{33} } \hfill & {M_{34} } \hfill \\
 {M_{41} } \hfill & {M_{42} } \hfill \ {M_{43} } \hfill & {M_{44} } \hfill \\
\end{array} }} \right)
\left( {{\begin{array}{*{20}c}
 {S_{0} } \hfill \\ {S_{1} } \hfill \\  {S_{2} } \hfill \\  {S_{3} } \hfill \\
\end{array} }} \right)_{\mathrm{input}},
\end{equation}
with Stokes vector components defined here by $S_{0}=I_{p}+I_{s}$, $S_{1}=I_{p} - I_{s}$, $S_{2}=I_{45}-I_{ -45}$, $S_{3}=I_{\sigma + }-I_{\sigma - }$. Here, $I_{p}$, $I_{s}$, $ I_{45}$, $I_{-45}$, $I_{\sigma + }$, and $I_{\sigma - }$denote the intensities for the $p$-, $s$-, +45$^{\circ}$, -45$^{\circ}$, right handed, and left handed circularly polarized light components, respectively.\cite{SchubertIRSEBook_2004} 

For our $c$-plane sample, a simpler representation is chosen as anisotropy is not observable. In this representation, we measure the change in polarization of light ($\tilde{\rho}$) caused by interaction with the sample. In reflection ellipsometry this change is described by
\begin{equation}
    \tilde{\rho} = \frac{\tilde{r}_{p}}{\tilde{r}_{s}} = \tan(\Psi)e^{i\Delta},
\end{equation}
where $\tilde{r}_{p}$ and $\tilde{r}_{s}$ are the Fresnel reflection coefficients for the light polarized parallel (\emph{p}) to and perpendicular (\emph{s}) to the plane of incidence. The rotation of the light's polarization state about the axis of propagation is defined as $\Psi$ and the relative phase shift between the parallel and perpendicular components is defined as $\Delta$. Here we report a $\Psi$, $\Delta$ pair for each wavenumber.

GSE measurements were taken in ambient conditions on two different instruments. Infrared data (450~\rzcm - 1200~\rzcm) was measured on a commercially available variable angle of incidence spectroscopic ellipsometer (VASE) (IR-VASE Mark-II; J.~A.~Woollam~Co.,~Inc.). Far-IR data (100~\rzcm - 450~\rzcm) was measured on an in-house built FIR-VASE system.\cite{KuehneRSI2014,8331870} For these measurements, two angles of incidence were selected as $\Phi_{a}$ = 50$^{\circ}$ and 70$^{\circ}$. For the $m$-plane sample, four azimuthal rotations (0$^{\circ}$, 45$^{\circ}$, 90$^{\circ}$, 135$^{\circ}$) were measured to allow us to fully observe the uniaxial optical symmetry. All data on the $m$-plane sample was taken using the Mueller matrix formalism, while the $c$-plane measurements utilized $\Psi$ and $\Delta$ notation. We note that only the IR ellipsometer instrument is capable of measuring the fourth row elements (excluding M$_{44}$) of the Mueller matrix, hence these are only available above approximately 250~\rzcm here.

Analysis of this data was completed using WVASE32$^{\mathrm{TM}}$ (J.~A.~Woollam Co.,~Inc.). In this uniaxial optical symmetry, group theory predicts two TO-LO mode pairs along the extraordinary direction defined here as $\varepsilon_{\parallel}$. Along the ordinary direction, $\varepsilon_{\bot}$, four TO-LO pairs are predicted. It has been established that in the IR spectral region nanoscale roughness has negligible effect on the observed experimental data\cite{SchubertIRSEBook_2004} and as a result, is not included in our model. We model the dielectric function tensor of the films directly against the experimental data using a Lorentzian oscillator model. Specifically, we utilize the four-parameter semi-quantum (FPSQ) model first described by Gervais and Periou.\cite{Gervais74} This model enables us to directly fit for both TO and LO mode frequency and broadening parameters independently ( $\omega_{\mathrm{TO,}l,j}$, $\gamma_{\mathrm{TO},l,j}$, $\omega_{\mathrm{LO,}l,j}$, $\gamma_{\mathrm{LO},l,j}$). The resultant dielectric function is hence modeled using
\begin{equation}\label{eq:4PSQmodel}
    \varepsilon_{j} = \varepsilon_{\infty,j} \prod_{l=1}^{N} \frac{\omega^{2}_{\mathrm{LO,}l,j}-\omega^{2}-i\omega\gamma_{\mathrm{LO,}l,j}}{\omega^{2}_{\mathrm{TO,}l,j}-\omega^{2}-i\omega\gamma_{\mathrm{TO,}l,j}},
\end{equation}
where $\varepsilon_{\infty,j}$ is the high-frequency dielectric constant. Here, $j$ denotes the symmetry (either $j = \parallel$ or $j = \bot$) and N denotes the number of phonon mode pairs in either direction. From this model, we can then derive the Lydanne-Sachs-Teller (LST) relationship\cite{Lyddane41}
\begin{equation}
    \frac{\varepsilon_{\mathrm{DC,}j}}{\varepsilon_{\infty,j}} = \prod_{l=1}^{N} \frac{\omega_{\mathrm{LO},l,j}^{2}}{\omega_{\mathrm{TO},l,j}^{2}},
\end{equation}
that leads to the determination of the static dielectric constant $\varepsilon_{\mathrm{DC,}j}$ in either direction.

\section{Results}

\subsection{Phonon mode displacement pattern}

\begin{figure}%[hbt]
\centering
\includegraphics[width=.9\linewidth]{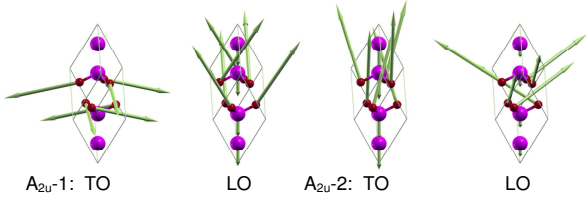}
\caption{\label{fig:a2u_modes} Mode displacement pattern for IR-active transverse (TO) and longitudinal (LO) modes with $A_{\mathrm{2u}}$ symmetry.}
\end{figure}
\begin{figure}%[hbt]
\centering
\includegraphics[width=.9\linewidth]{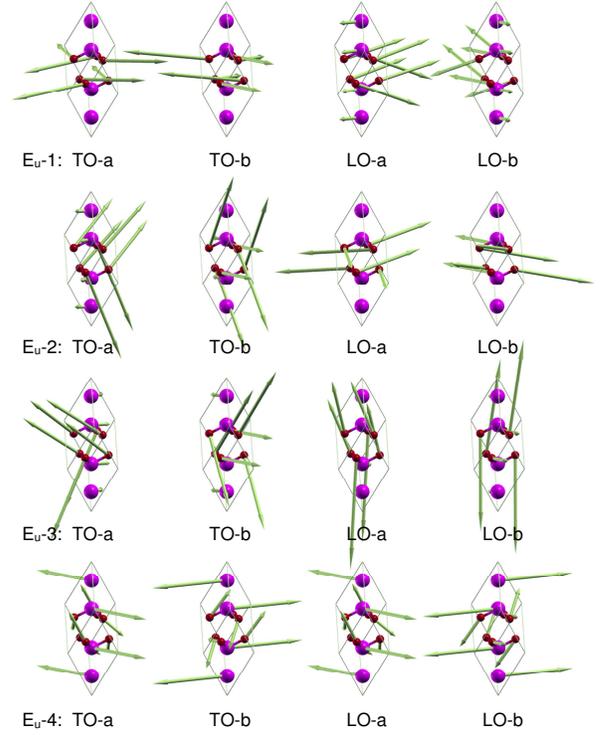}
\caption{\label{fig:eu_modes} Same as Figure~\ref{fig:a2u_modes} for $E_{\mathrm{u}}$ symmetry.}
\end{figure}
\begin{figure}%[hbt]
\centering
\includegraphics[width=.9\linewidth]{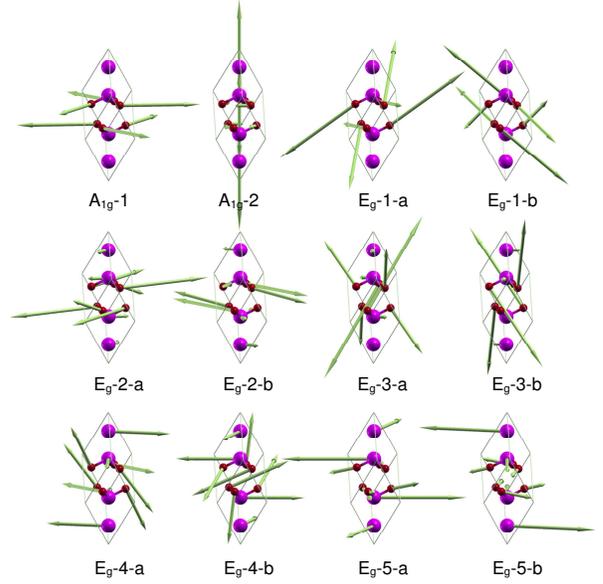}
\caption{\label{fig:raman_modes} Mode displacement pattern for Raman-active modes with $A_{\mathrm{1g}}$ and $E_{\mathrm{g}}$ symmetry.}
\end{figure}
\begin{figure}%[hbt]
\centering
\includegraphics[width=.675\linewidth]{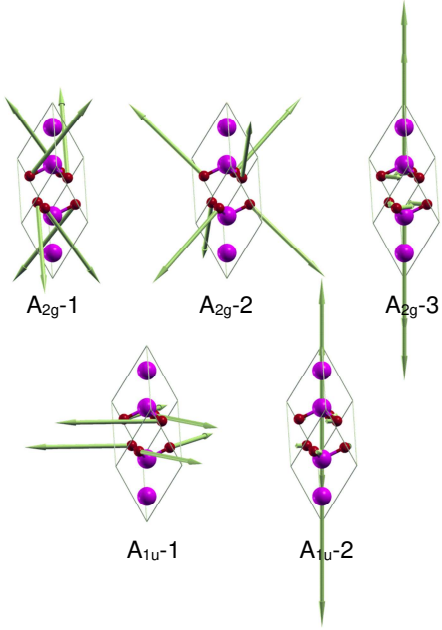}
\caption{\label{fig:silent_modes} Mode displacement pattern for silent modes with $A_{\mathrm{2g}}$ and $A_{\mathrm{1u}}$ symmetries. DFT-calculated LDA and GGA frequencies, respectively, of these modes are: 705.2 and 671.2 cm$^{-1}$ ($A_{\mathrm{2g}}$-1), 483.2 and 465.2 cm$^{-1}$ ($A_{\mathrm{2g}}$-2), 164.9 and 150.1 cm$^{-1}$ ($A_{\mathrm{2g}}$-3), 627.7 and 599.4 cm$^{-1}$ ($A_{\mathrm{1u}}$-1), and 346.7 and 328.5 cm$^{-1}$ ($A_{\mathrm{1u}}$-2).}
\end{figure}

Figures~\ref{fig:a2u_modes}--~\ref{fig:silent_modes} depict the atomic displacement pattern for all zone-center phonon modes discussed in this work. Note that these pattern establish the eigenvectors of the dynamic matrix, which populate a 3$N$ dimensional space, where $N=10$ is the number of nucleus within the unit cell of $\alpha$-Ga$_{2}$O$_3$. We will further below use these eigenvetors to perform similarity comparisons between the isotructural compounds $\alpha$-Ga$_{2}$O$_3$ and $\alpha$-Al$_{2}$O$_3$. Note further that all displacement eigenvectors within the same symmetry group, and within the same compound, are orthogonal to each other. For example, the eigenvector belonging to $E_{\mathrm{u}}-1$ TO in $\alpha$-Ga$_{2}$O$_3$ is orthogonal to any of the other modes $E_{\mathrm{u}}-2,3,4$ in $\alpha$-Ga$_{2}$O$_3$, etc. However, this requirement is relaxed when comparing the same vectors between $\alpha$-Ga$_{2}$O$_3$ and $\alpha$-Al$_{2}$O$_3$, as will be done further below.

\subsection{Phonon mode eigenvector overlaps}

\begin{table}\centering
\caption{Eigenvector overlaps for $A_{\mathrm{1g}}$ Raman-active modes between aGO and aAO.}
\label{tab:ag_overlaps}
\begin{ruledtabular}
\begin{tabular}{{c}{c}{c}{c}}
& & \multicolumn{2}{c}{aAO} \\
\hline
& & $A_{\mathrm{1g}}$-1 & $A_{\mathrm{1g}}$-2\\
\hline
\multirow{2}{*}{aGO}&$A_{\mathrm{1g}}$-1 & 0.99970 & 0.02465\\ 
&$A_{\mathrm{1g}}$-2 & 0.02465 & 0.99970\\ 
\end{tabular}
\end{ruledtabular}
\end{table}

\begin{table}\centering
\caption{Same as Table~\ref{tab:ag_overlaps} for $E_{\mathrm{g}}$ Raman-active modes.}
\label{tab:eg_overlaps}
\begin{ruledtabular}
\begin{tabular}{{c}{c}{c}{c}{c}{c}{c}}
& & \multicolumn{5}{c}{aAO} \\
%\cline{8-12}
\hline
& & $E_{\mathrm{g}}$-1 & $E_{\mathrm{g}}$-2 & $E_{\mathrm{g}}$-3 & $E_{\mathrm{g}}$-4 & $E_{\mathrm{g}}$-5\\
\hline
&$E_{\mathrm{g}}$-1 & 0.99830 &	0.01711 & 0.03720 & 0.01792 & 0.03753\\
&$E_{\mathrm{g}}$-2 & 0.01317 & 0.76171 & 0.37537 & 0.51318 & 0.12396\\ 
aGO&$E_{\mathrm{g}}$-3 & 0.02542 & 0.28638 & 0.33607 & 0.03656 & 0.89614\\ 
&$E_{\mathrm{g}}$-4 & 0.01050 &	0.56935 & 0.47976 &	0.57555 & 0.33809\\ 
&$E_{\mathrm{g}}$-5 & 0.04976 & 0.11547 & 0.71736 & 0.63540 & 0.25663\\ 
%\hline
%&$E_{\mathrm{g}}$-1 & & & & & & 1.00 & 0.00 & 0.00 & 0.00 & 0.00\\ 
%&$E_{\mathrm{g}}$-2 & & & & & & & 1.00 & 0.00 & 0.00 & 0.00\\ 
%aAO&$E_{\mathrm{g}}$-3 & & & & & & & & 1.00 & 0.00 & 0.00\\ 
%&$E_{\mathrm{g}}$-4 & & & & & & & & & 1.00 & 0.00\\ 
%&$E_{\mathrm{g}}$-5 & & & & & & & & & & 1.00\\ 
\end{tabular}
\end{ruledtabular}
\end{table}

\begin{table}\centering
\caption{Eigenvector overlaps for $A_{\mathrm{2g}}$ silent modes.}
\label{tab:a2g_overlaps}
\begin{ruledtabular}
\begin{tabular}{{c}{c}{c}{c}{c}}
& & \multicolumn{3}{c}{aAO} \\
\hline
& & $A_{\mathrm{2g}}$-1 & $A_{\mathrm{2g}}$-2 & $A_{\mathrm{2g}}$-3\\
\hline
\multirow{3}{*}{aGO}&$A_{\mathrm{2g}}$-1 & 0.99312 & 0.09770 & 0.06455\\ 
&$A_{\mathrm{2g}}$-2 & 0.09679 & 0.99516 & 0.01705\\ 
&$A_{\mathrm{2g}}$-3 & 0.06590 & 0.01068 & 0.99777\\ 
\end{tabular}
\end{ruledtabular}
\end{table}

\begin{table}\centering
\caption{Same as Table~\ref{tab:ag_overlaps} for $A_{\mathrm{1u}}$ silent modes.}
\label{tab:a1u_overlaps}
\begin{ruledtabular}
\begin{tabular}{{c}{c}{c}{c}}
& & \multicolumn{2}{c}{aAO} \\
\hline
& & $A_{\mathrm{1u}}$-1 & $A_{\mathrm{1u}}$-2\\
\hline
\multirow{2}{*}{aGO}&$A_{\mathrm{1u}}$-1 & 0.93818 & 0.34616\\ 
&$A_{\mathrm{1u}}$-2 & 0.34616 & 0.93818\\ 
\end{tabular}
\end{ruledtabular}
\end{table}

\begin{table*}\centering
\caption{Eigenvector overlaps for IR-active $A_{\mathrm{2u}}$ TO and LO modes.}
\label{tab:a2u_overlaps}
\begin{ruledtabular}
\begin{tabular}{{l}{c}{c}{c}{c}{c}{c}{c}{c}{c}}
& & \multicolumn{4}{c}{aGO} & \multicolumn{4}{c}{aAO} \\
\cline{3-6}\cline{7-10}
& & $A_{\mathrm{2u}}$-1 (TO) & $A_{\mathrm{2u}}$-2 (TO) & $A_{\mathrm{2u}}$-1 (LO) & $A_{\mathrm{2u}}$-2 (LO) & $A_{\mathrm{2u}}$-1 (TO) & $A_{\mathrm{2u}}$-2 (TO) & $A_{\mathrm{2u}}$-1 (LO) & $A_{\mathrm{2u}}$-2 (LO)\\
\hline
\multirow{4}{*}{aGO}&$A_{\mathrm{2u}}$-1 (TO) & 1.0 & 0.0 & 0.74439 & 0.66774 & 0.98123 & 0.18222 & 0.54780 & 0.83423\\ 
&$A_{\mathrm{2u}}$-2 (TO) & & 1.0 & 0.66774 & 0.74439 & 0.16497 & 0.96268 & 0.80890 & 0.54740\\ 
&$A_{\mathrm{2u}}$-1 (LO) & & & 1.0 & 0.0 & 0.84058 & 0.50718 & 0.94791 & 0.25547\\ 
&$A_{\mathrm{2u}}$-2 (LO) & & & & 1.0 & 0.53240 & 0.83829 & 0.23635 & 0.96453\\ 
\hline
\multirow{4}{*}{aAO}&$A_{\mathrm{2u}}$-1 (TO) & & & & & 1.0 & 0.0 & 0.69229 & 0.72161\\ 
&$A_{\mathrm{2u}}$-2 (TO) & & & & & & 1.0 & 0.72162 & 0.69229\\ 
&$A_{\mathrm{2u}}$-1 (LO) & & & & & & & 1.0 & 0.0\\ 
&$A_{\mathrm{2u}}$-2 (LO) & & & & & & & & 1.0\\ 
\end{tabular}
\end{ruledtabular}
\end{table*}

\begin{sidewaystable*}[]\centering
\caption{Same as Table~\ref{tab:a2u_overlaps} for IR-active $E_{\mathrm{u}}$ TO and LO modes.}
\label{tab:eu_overlaps}
\begin{ruledtabular}
\begin{tabular}{{l}{c}{c}{c}{c}{c}{c}{c}{c}{c}{c}{c}{c}{c}{c}{c}{c}{c}}
& & \multicolumn{8}{c}{aGO} & \multicolumn{8}{c}{aAO} \\
\cline{3-10}\cline{11-18}
& & \multicolumn{4}{c}{TO modes} & \multicolumn{4}{c}{LO modes} & \multicolumn{4}{c}{TO modes} & \multicolumn{4}{c}{LO modes} \\
\cline{3-6} \cline{7-10} \cline{11-14} \cline{15-18}
& & $E_{\mathrm{u}}$-1 & $E_{\mathrm{u}}$-2 & $E_{\mathrm{u}}$-3 & $E_{\mathrm{u}}$-4 & $E_{\mathrm{u}}$-1 & $E_{\mathrm{u}}$-2 & $E_{\mathrm{u}}$-3 & $E_{\mathrm{u}}$-4 & $E_{\mathrm{u}}$-1 & $E_{\mathrm{u}}$-2 & $E_{\mathrm{u}}$-3 & $E_{\mathrm{u}}$-4 & $E_{\mathrm{u}}$-1 & $E_{\mathrm{u}}$-2 & $E_{\mathrm{u}}$-3 & $E_{\mathrm{u}}$-4\\
\hline
&$E_{\mathrm{u}}$-1 & 1.0 & 0.0 & 0.0 & 0.0 & 0.17661 &	0.98387 & 0.02830 & 0.00026 & 0.99430 &	0.09687 & 0.00179 & 0.03035 & 0.14752 &	0.98788 & 0.01486 & 0.03254\\ 
aGO&$E_{\mathrm{u}}$-2 & & 1.0 & 0.0 & 0.0 & 0.81323 & 0.16206 & 0.55891 & 0.00318 & 0.08513 & 0.97046 & 0.09036 & 0.14550 & 0.75749 & 0.13280 & 0.59598 & 0.17822\\ 
(TO)&$E_{\mathrm{u}}$-3 & & & 1.0 & 0.0 & 0.55445 & 0.07569 & 0.82873 & 0.00785 & 0.01824 & 0.11243 & 0.91013 & 0.36243 & 0.59653 & 0.07595 & 0.74964 & 0.22156\\ 
&$E_{\mathrm{u}}$-4 & & & & 1.0 & 0.00698 & 0.00085 & 0.00472 & 0.99996 & 0.03699 & 0.09908 & 0.37925 & 0.91923 & 0.00549 & 0.02641 & 0.28512 & 0.95811\\ 
&$E_{\mathrm{u}}$-1 & & & & & 1.0 & 0.0 & 0.0 & 0.0 & 0.25469 & 0.83375 & 0.43347 & 0.07085 & 0.97277 & 0.02418 & 0.06966 & 0.03452\\ 
aGO&$E_{\mathrm{u}}$-2 & & & & & & 1.0 & 0.0 & 0.0 & 0.96312 & 0.26100 & 0.05633 & 0.03293 & 0.02277 & 0.99924 & 0.02497 & 0.01909\\ 
(LO)&$E_{\mathrm{u}}$-3 & & & & & & & 1.0 & 0.0 & 0.06077 & 0.44696 & 0.80659 & 0.37820 & 0.06679 & 0.01680 & 0.95612 & 0.27962\\ 
&$E_{\mathrm{u}}$-4 & & & & & & & & 1.0 & 0.03766 & 0.10302 & 0.37239 & 0.92157 & 0.01261 & 0.02565 & 0.28113 & 0.95924\\ 
\hline
&$E_{\mathrm{u}}$-1 & & & & & & & & & 1.0 & 0.0 & 0.0 & 0.0 & 0.23308 & 0.97054 & 0.06059 & 0.00751\\ 
aAO&$E_{\mathrm{u}}$-2 & & & & & & & & & & 1.0 & 0.0 & 0.0 & 0.82422 & 0.23047 & 0.51499 & 0.04821\\ 
(TO)&$E_{\mathrm{u}}$-3 & & & & & & & & & & & 1.0 & 0.0 & 0.50303 & 0.06889 & 0.84939 & 0.14405\\ 
&$E_{\mathrm{u}}$-4 & & & & & & & & & & & & 1.0 & 0.11529 & 0.01391 & 0.09821 & 0.98837\\ 
&$E_{\mathrm{u}}$-1 & & & & & & & & & & & & & 1.0 & 0.0 & 0.0 & 0.0\\ 
aAO&$E_{\mathrm{u}}$-2 & & & & & & & & & & & & & & 1.0 & 0.0 & 0.0\\ 
(LO)&$E_{\mathrm{u}}$-3 & & & & & & & & & & & & & & & 1.0 & 0.0\\ 
&$E_{\mathrm{u}}$-4 & & & & & & & & & & & & & & & & 1.0\\ 
\end{tabular}
\end{ruledtabular}
\end{sidewaystable*}

In order to quantify the comparison of the phonon modes between aGO and aAO, we calculate the overlaps between the eigenvectors which represent the atomic displacement patterns presented in Figs.~\ref{fig:a2u_modes}-\ref{fig:silent_modes}. The overlaps are often used to compare phonon modes, although the specific details of how results are obtained can differ.\cite{Zicovich-Wilson_2008,Lyu_2019} In our work here, the overlaps are calculated as dot products between the respective eigenvectors in the 3$N$ dimensional vector space in which the eigenvectors are established, where $N$ is the number of atoms in the primitive crystal cell, and divided by their norms. Hence, the overlaps can be thought as projections of one eigenvector onto another. It is worth noting that by definition the eigenvectors are orthogonal, i.e., the dot product of an eigenvector with another eigenvector from the same set of eigenvectors (eigenvectors of the same matrix) is always zero. Also, the dot products of eigenvectors belonging to different irreducible representations are always zero. The overlaps we report here are normalized. This can be seen in the tables, where the sum of the overlaps across all eigenvectors within the same irreducible representation, when each as shown is squared and then summed up, along each row or each column, add up to 1 within the numerical uncertainty.\footnote{Note that overlaps, in principle, can be negative if the eigenvectors point in opposite directions. As we compare eigenvectors obtained from the diagonalization of independent dynamic matrices this is not unusual, but physically insignificant and we ignore the sign.} In the case of IR-active modes the situation is slightly more complicated: the IR-active modes may exist as either TO or LO modes. We report them all in the same table, but the TO modes and the LO modes are obtained independently as eigenvectors of different dynamical matrices and need to be considered separately. Hence, when we for example look at mode $A_{2\mathrm{u}}-1$ (TO) for aGO, along the top row of Table~\ref{tab:a2u_overlaps}, the normalization can be observed four times, in each pair of columns representing a consistent pair of eigenvectors: TO modes for aGO (simple orthogonality of eigenvectors), LO modes for aGO, TO modes for aAO, and LO modes for aAO. Another feature specific to the IR-active modes is that the three acoustic modes (and their eigenvectors) belong to the same irreducible representations as the IR-active optical modes (see section~\ref{structure}).  Because we do not include the acoustic eigenmode overlap values in the tables, sums of the squares of IR-active mode overlaps are slightly less than 1.

Tables~\ref{tab:ag_overlaps} and~\ref{tab:eg_overlaps} list the eigenvector overlaps for Raman-active $A_{\mathrm{1g}}$ and $E_{\mathrm{g}}$ modes, respectively. Tables~\ref{tab:a2g_overlaps} and~\ref{tab:a1u_overlaps} summarize the overlaps for the silent modes $A_{\mathrm{2g}}$ and $A_{\mathrm{1u}}$, respectively. Tables~\ref{tab:a2u_overlaps} and~\ref{tab:eu_overlaps} expand the overlaps for TO and LO modes with $A_{\mathrm{2u}}$ and $E_{\mathrm{u}}$ symmetries, respectively.\footnote{Note that the dot product is a symmetric operation, hence, Tables~\ref{tab:a2u_overlaps} and~\ref{tab:eu_overlaps} are symmetric, and only the upper portions are shown.  Tables~\ref{tab:ag_overlaps},~\ref{tab:eg_overlaps},~\ref{tab:a2g_overlaps}, and ~\ref{tab:a1u_overlaps} are written in a shortened manner and are thus not symmetric because the dot product of mode vector $A$ in compound X with mode vector $B$ in compound Y differs from the dot product of vector $B$ in compound X with vector $A$ in compound Y.}
\subsection{Phonon dispersion}
\begin{figure}%[hbt]
\centering
\includegraphics[width=.80\linewidth]{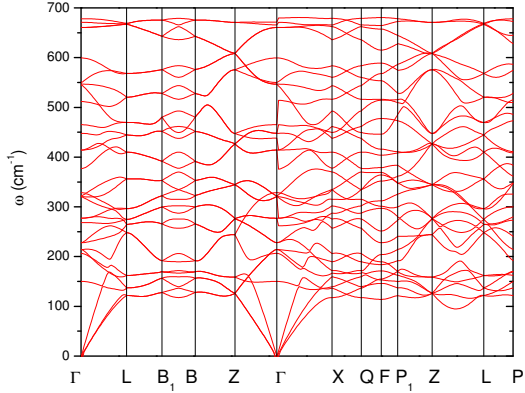}
\caption{\label{fig:phd} Phonon dispersion in aGO. Notation as in Ref.~\onlinecite{setyawan2010} (Brillouin zone for the type-1 rhombohedral lattice, RHL$_1$).}
\end{figure}

Figure~\ref{fig:phd} depicts the phonon dispersion of all modes along a high symmetry path in the Brillouin zone obtained using the DFT-GGA calculations with symmetry notation as in Ref.~\onlinecite{setyawan2010}. Qualitatively, the phonon dispersions for aGO are quite similar to those of aAO (see, for example, Ref.~\onlinecite{mp1143}). The generally larger LO-TO splitting for aAO is evident in the larger separation of frequencies at the $\Gamma$ point, particularly for the modes on the high-frequency end of the spectrum.

\subsection{Raman-active and silent phonon mode parameters}

\begin{table*}\centering 
\caption{\label{Table:Raman}Parameters for $\alpha$-Ga$_{2}$O$_3$ Raman-active phonon modes obtained from DFT calculations in this work, and compared with available literature data. Included are Raman frequencies $\omega_{\mathrm{RA}}$ and Raman scattering activities $S_{\mathrm{RA}}$. Note that due to limitations of the algorithm for calculating the second order response in Quantum ESPRESSO (and similar codes) the calculation of Raman scattering activity at the GGA level of theory only includes the local part of the density functional without the gradient correction.}
\begin{ruledtabular}
\begin{tabular}{{l}{c}{c}{c}{c}{c}{c}{c}{c}}  
Mode &  $\omega_{\mathrm{RA}}$ & $\omega_{\mathrm{RA}}$ & $S_{\mathrm{RA}}$ & $S_{\mathrm{RA}}$ & $\omega_{\mathrm{RA}}$ & $\omega_{\mathrm{RA}}$& $\omega_{\mathrm{RA}}$& $\omega_{\mathrm{RA}}$\\
& This work & This work  & This work  & This work &  Ref.~\onlinecite{Cusco15} & Ref.~\onlinecite{Feneberg19aGO} &  Ref.~\onlinecite{Cusco15} & Ref.~\onlinecite{Sharma2021}\\
& LDA & GGA & LDA & GGA & Exp. & Exp. & LDA & LDA\\
& (cm$^{-1}$) & (cm$^{-1}$) & $\left(\frac{\mathrm{\AA}^4}{\mathrm{amu}}\right)$ & $\left(\frac{\mathrm{\AA}^4}{\mathrm{amu}}\right)$ & (cm$^{-1}$) & (cm$^{-1}$)& (cm$^{-1}$)& (cm$^{-1}$)\\\hline
$A_{\mathrm{1g}}$-1 & 588.27 & 550.1 & 9.50 & 14.9 & 569.7 & 574.3 & 551& 555.1\\
$A_{\mathrm{1g}}$-2 & 216.77 & 206.5 & 1.36 & 1.75 & 218.2 & 218.0 & 211& 210.3\\
\hline
$E_{\mathrm{g}}$-1 & 708.70 & 660.6 & 4.67 & 6.43 & 686.7 & 688.4 & 667& 669.1 \\
$E_{\mathrm{g}}$-2 & 458.26 & 414.1 & 0.909 & 1.46 & 430.7 & 431.3 & 426& 426.7 \\
$E_{\mathrm{g}}$-3 & 336.15 & 320.4 & 0.138 & 0.233 & 328.8 & 328.0 & 314&315.3 \\
$E_{\mathrm{g}}$-4 & 299.03 & 277.2 & 0.552 & 1.16 & 285.3 & 286.8 & 281&282.6\\
$E_{\mathrm{g}}$-5 & 236.44 & 227.9 & 0.0365 & 0.0248 & 240.7 & 241.9 & 233&236.6\\
\end{tabular}
\end{ruledtabular}
\end{table*}

\begin{table}\centering
\caption{\label{Table:DFT_AORaman} Parameters for $\alpha$-Al$_2$O$_3$ Raman-active phonon modes obtained from DFT calculations at the GGA level in this work. Included are Raman frequencies $\omega_{\mathrm{RA}}$ and Raman scattering activities $S_{\mathrm{RA}}$. Note that due to limitations of the algorithm for calculating the second order response in Quantum ESPRESSO (and similar codes) the calculation of Raman scattering activity at the GGA level of theory only includes the local part of the density functional without the gradient correction.}
\begin{ruledtabular}
\begin{tabular}{{l}{c}{c}}
Mode & $\omega_{\mathrm{RA}}$  & $S_{\mathrm{RA}}$\\
& (cm$^{-1}$)  &  $\left(\frac{\mathrm{\AA}^4}{\mathrm{amu}}\right)$\\
\hline
$A_{\mathrm{1g}}$-1 & 609.2 & 0.995\\
$A_{\mathrm{1g}}$-2 & 390.8 & 1.95\\
\hline
$E_{\mathrm{g}}$-1 & 714.0 & 1.01\\
$E_{\mathrm{g}}$-2 & 547.2 & 0.109\\
$E_{\mathrm{g}}$-3 & 423.1 & 0.0525\\
$E_{\mathrm{g}}$-4 & 401.5 & 0.150\\
$E_{\mathrm{g}}$-5 & 358.4 & 0.266\\
\end{tabular}
\end{ruledtabular}
\end{table}

The parameters for $\alpha$-Ga$_2$O$_3$ Raman-active modes with $A_{\mathrm{1g}}$ and $E_{\mathrm{g}}$ symmetries are collected in Table~\ref{Table:Raman}, and are compared with available theoretical and experimental values. Data are reported from DFT calculations using LDA and GGA obtained in this  work, from experimental investigations by Cuscó~\textit{et al.}~\cite{Cusco15} and Feneberg~\textit{et al.},~\cite{Feneberg19aGO} and DFT calculations using LDA by Cuscó~\textit{et al.}~\cite{Cusco15} and Sharma and Singisetti.~\cite{Sharma2021} DFT calculated parameters for the silent modes with $A_{\mathrm{2g}}$ and $A_{\mathrm{1u}}$ symmetries are given in the caption of Figure~\ref{fig:silent_modes}. These modes have not been experimentally observed so far. We include them here for the sake of completeness. Parameters for $\alpha$-Al$_2$O$_3$ Raman-active phonon modes obtained from DFT calculations at the GGA level in this work are shown in Table~\ref{Table:DFT_AORaman} for comparison.

\subsection{Infrared-active phonon mode parameters}

\subsubsection{DFT}

\begin{table*}\centering 
\caption{\label{Table:DFT}Parameters for $\alpha$-Ga$_2$O$_3$ IR-active phonon modes obtained from DFT calculations in this work, and compared with available literature data.}
\begin{ruledtabular}
\begin{tabular}{{l}{c}{c}{c}{c}{c}{c}{c}{c}{c}{c}{c}}  
%& \multicolumn{11}{c}{$\alpha$-Ga$_2$O$_3$} & \multicolumn{4}{c}{$\alpha$-Al$_2$O$_3$} \\ %\cline{2-12}\cline{13-16}
Mode & $\omega_{\mathrm{TO}}$ & $\omega_{\mathrm{TO}}$& $\omega_{\mathrm{TO}}$& $\omega_{\mathrm{TO}}$ & $\omega_{\mathrm{LO}}$ & $\omega_{\mathrm{LO}}$ & $\omega_{\mathrm{LO}}$& $A^2_{\mathrm{TO}}$& $A^2_{\mathrm{TO}}$& $A^2_{\mathrm{LO}}$& $A^2_{\mathrm{LO}}$\\
&This work&This work&Ref.~\onlinecite{Feneberg18aGO}&Ref.~\onlinecite{Sharma2021}&This work&This work&Ref.~\onlinecite{Sharma2021}&This work&This work&This work&This work\\
&LDA&GGA&LDA&LDA&LDA&GGA&LDA&LDA&GGA&LDA&GGA\\
& (cm$^{-1}$) &  (cm$^{-1}$) & (cm$^{-1}$)  &  (cm$^{-1}$) & (cm$^{-1}$) &(cm$^{-1}$) &(cm$^{-1}$) & $\left[\frac{(\mathrm{D}/\mathrm{\AA})^2}{\mathrm{amu}}\right]$ & $\left[\frac{(\mathrm{D}/\mathrm{\AA})^2}{\mathrm{amu}}\right]$ & $\left[\frac{(\mathrm{D}/\mathrm{\AA})^2}{\mathrm{amu}}\right]$ & $\left[\frac{(\mathrm{D}/\mathrm{\AA})^2}{\mathrm{amu}}\right]$\\
\hline
$A_{\mathrm{2u}}$-1 & 546.65 & 514.52 & 531 & 535.1 & 705.47 & 665.92 & 674.27& 15.03 & 12.39 & 48.45 & 49.66\\
$A_{\mathrm{2u}}$-2 & 292.81 & 261.35 & 269 & 270.6 & 453.43 & 437.82 &435.29 & 40.03 & 43.94 & 6.609 & 6.676\\
\hline
$E_{\mathrm{u}}$-1 & 584.23 & 546.65 & 551& 553.6 & 719.27 & 679.38& 682.18& 1.019 & 0.5493 & 53.03 & 55.07\\
$E_{\mathrm{u}}$-2 & 494.05 & 447.48 & 462&462.2 & 580.84 & 544.68 &546.19& 30.11 & 30.03 & 0.4510 & 0.2978\\
$E_{\mathrm{u}}$-3 & 353.59 & 322.72 & 326 &328.8 & 408.57 & 375.83 &373.51 & 24.02 &26.11 & 1.673 & 1.326\\
$E_{\mathrm{u}}$-4 & 229.53 & 213.97 & 220 &220.8 & 229.55 & 213.99 &219.94& 0.0035 & 0.0056 & 0.0002 & 0.0003\\
\end{tabular}
\end{ruledtabular}
\end{table*}

\begin{table}\centering
\caption{\label{Table:DFT_AO} Parameters for $\alpha$-Al$_2$O$_3$ IR-active phonon modes obtained from  DFT calculations at the GGA level in this work.}
\begin{ruledtabular}
\begin{tabular}{{l}{c}{c}{c}{c}}
Mode & $\omega_{\mathrm{TO}}$  & $\omega_{\mathrm{LO}}$& $A^2_{\mathrm{TO}}$ & $A^2_{\mathrm{LO}}$\\
& (cm$^{-1}$)  &  (cm$^{-1}$) & $\left[\frac{(\mathrm{D}/\mathrm{\AA})^2}{\mathrm{amu}}\right]$ & $\left[\frac{(\mathrm{D}/\mathrm{\AA})^2}{\mathrm{amu}}\right]$\\
\hline
$A_{\mathrm{2u}}$-1 & 546.65 & 829.66 & 20.31 & 62.77 \\
$A_{\mathrm{2u}}$-2 & 369.52 & 484.68 & 44.30 & 1.8361 \\
\hline
$E_{\mathrm{u}}$-1 & 601.58 & 858.66 & 2.134 & 64.18 \\
$E_{\mathrm{u}}$-2 & 533.15 & 598.17 & 38.86 & 0.1323 \\
$E_{\mathrm{u}}$-3 & 411.87  & 453.81 & 22.72 & 0.7485 \\
$E_{\mathrm{u}}$-4 & 358.89 & 360.85 & 1.372 & 0.0254 \\
\end{tabular}
\end{ruledtabular}
\end{table}

Table~\ref{Table:DFT} summarizes all parameters for IR-active phonon modes in $\alpha$-Ga$_2$O$_3$ obtained from DFT calculations in this work both at the LDA and GGA levels, compared with available literature data at the LDA level from Feneberg~\textit{et al.}~\cite{Feneberg18aGO} and Sharma and Singisetti.~\cite{Sharma2021} Table~\ref{Table:DFT_AO} presents our DFT calculated parameters for IR-active phonon modes in $\alpha$-Al$_2$O$_3$ at the GGA level for further comparison.

\subsubsection{Ellipsometry}

\begin{figure}%[hbt]
\centering
\includegraphics[width=.9\linewidth]{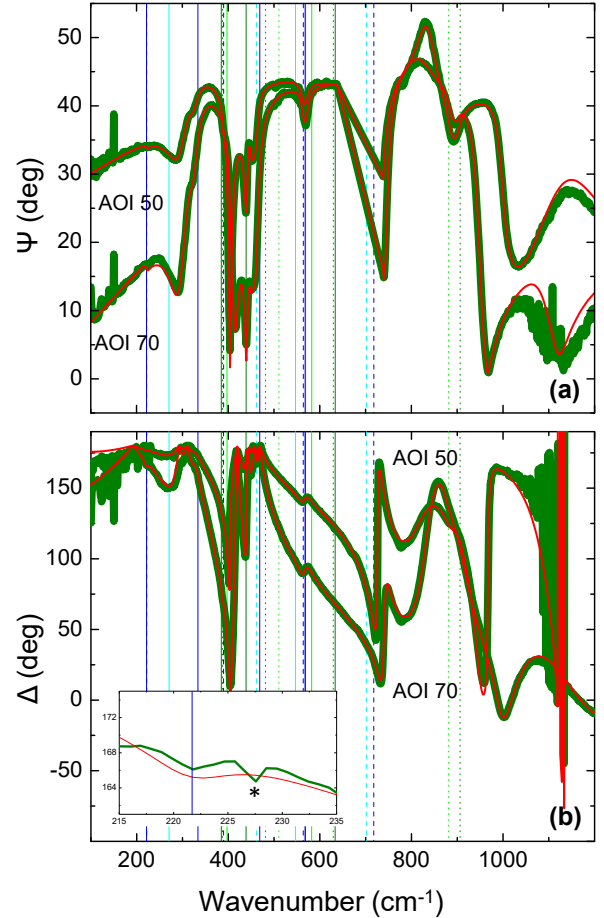}
\caption{\label{fig:psidel} Experimental (green dotted lines) and best match model calculated ellipsometry data $\Psi$: (a) and $\Delta$: (b) for the $\alpha$-Ga$_{2}$O$_3$ thin film sample grown on $c$-plane sapphire. Vertical lines indicate wavenumbers of IR-active modes (TO: solid lines, LO: dashed lines) for $\alpha$-Ga$_2$O$_3$ (Light blue: $A_{\mathrm{2u}}$; blue: $E_{\mathrm{u}}$) and $\alpha$-Al$_2$O$_3$ (Light green: $A_{\mathrm{2u}}$; green: $E_{\mathrm{u}}$), for comparison. Inset in pane (b) highlights data from 215-235~cm$^{-1}$ which shows the small feature associated with the lowest $E_{\mathrm{u}}$ mode.The asterisk denotes a small feature caused by water absorbance.}
\end{figure}

\begin{figure*}%[hbt]
\centering
\includegraphics[width=\linewidth]{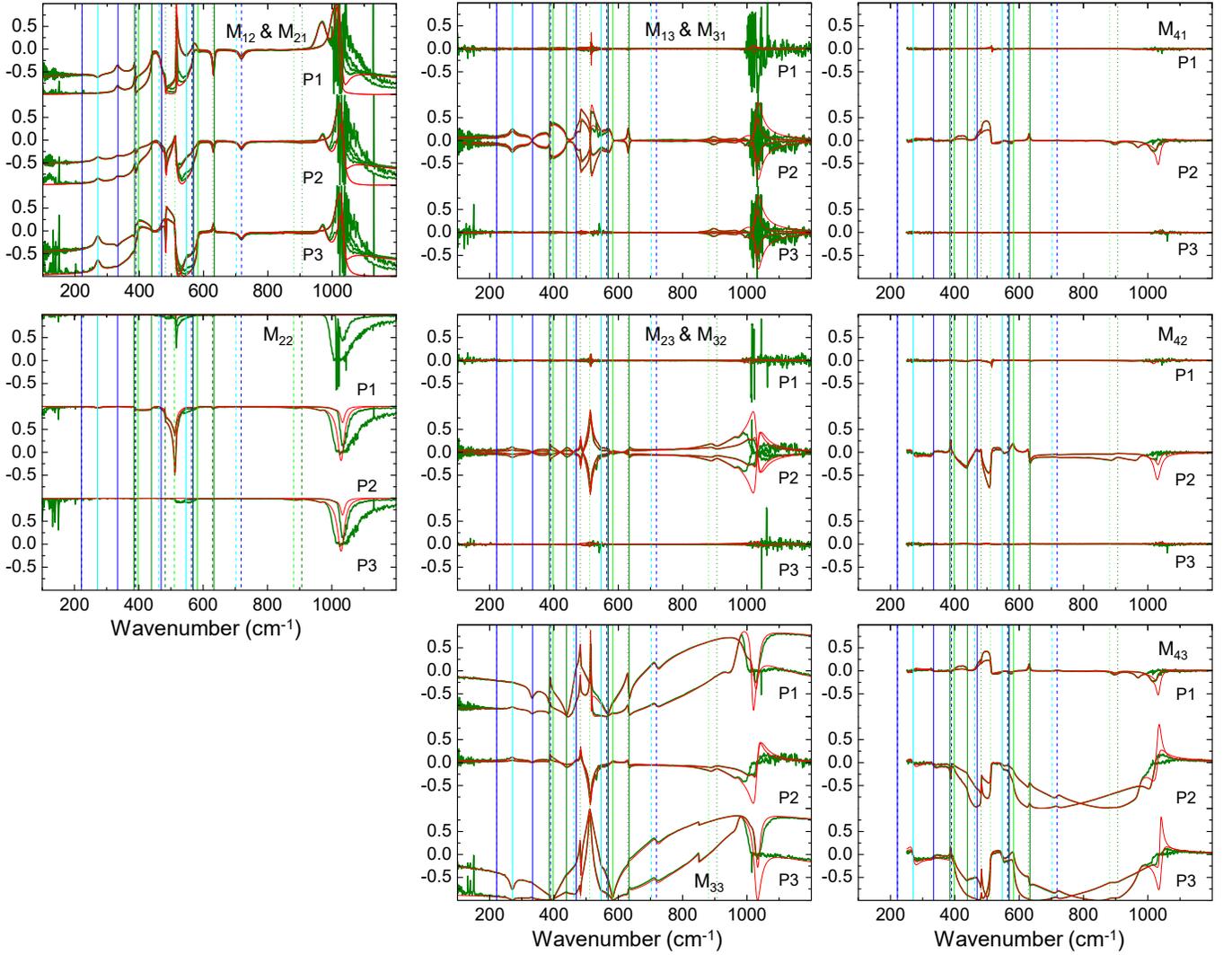}
\caption{\label{fig:MM} Same as Figure~\ref{fig:psidel} for ellipsometry Mueller matrix data for the thin film sample on $m$-plane sapphire. P1, P2, and P3 denote azimuthal sample rotation angles of 45$^{\circ}$, 90$^{\circ}$, and 135$^{\circ}$, respectively, relative to a side edge of the substrate with known crystallographic orientation. This information is used during data analysis.}
\end{figure*}

\begin{table*}\centering 
\caption{\label{Table:Ellip}FPSQ model parameters for $\alpha$-Ga$_2$O$_3$ TO and LO modes obtained from best match model dielectric function analysis. Same data for $\alpha$-Al$_2$O$_3$ determined previously by Schubert, Tiwald, and Herzinger are included for comparison (Ref.~\onlinecite{SchubertPRB61_2000}). Parameters denoted with an asterisk were range limited in the fit and hence have a relatively large uncertainty. All parameters are reported in units of cm$^{-1}$. Data for static and high frequency dielectric constants are summarized in Table~\ref{Table:epsilon}.}
\begin{ruledtabular}
\begin{tabular}{{l}{c}{c}{c}{c}{c}{c}{c}{c}{c}}
& \multicolumn{5}{c}{$\alpha$-Ga$_2$O$_3$} & \multicolumn{4}{c}{$\alpha$-Al$_2$O$_3$} \\ \cline{2-6}\cline{7-10}
 & \multicolumn{2}{c}{\scriptsize{$\omega_{\mathrm{TO}}$}}   & \multicolumn{1}{l}{} & \multicolumn{1}{l}{} \\\cline{2-3}
\scriptsize{Mode}& \scriptsize{This work} & \scriptsize{Ref.~\onlinecite{Feneberg19aGO}} &\scriptsize{$\omega_{\mathrm{LO}}$} &\scriptsize{$\gamma_{\mathrm{TO}}$}&\scriptsize{$\gamma_{\mathrm{LO}}$} & \scriptsize{$\omega_{\mathrm{TO}}$} &\scriptsize{$\omega_{\mathrm{LO}}$} &\scriptsize{$\gamma_{\mathrm{TO}}$}&\scriptsize{$\gamma_{\mathrm{LO}}$}\\
\hline
$A_{\mathrm{2u}}$-1 & 547.1 ± 0.6 & 546.6 &  702.3 ± 0.1 & 15.2 ± 1.1 & 19.7 ± 0.3&582.41  &  881.1 & 3.0  & 15.4\\
$A_{\mathrm{2u}}$-2  & 270.8 ± 0.1 & 271.3 & 461.8 ± 0.1 & 14.9 ± 0.2 &12.2 ± 0.2 & 397.5 &  510.9 & 5.3  & 1.1 \\ 
\hline
$E_{\mathrm{u}}$-1  & 568.5 ± 1.0 & 571.7 &  718.3 ± 0.4 & 16.9 ± 2.1 & 17.4 ± 0.9 & 633.6   &  906.6 & 5.0 & 14.7 \\
$E_{\mathrm{u}}$-2  & 469.5 ± 0.1 & 474.1 &  564.3 ± 1.0 & 19.8 ± 0.2 & 15.7 ± 2.1 & 569.0 &    629.5 & 4.7  & 5.9 \\
$E_{\mathrm{u}}$-3  & 334.0 ± 0.1 & 333.2 &  390.0 ± 0.1 & 16.4 ± 0.1 & 11.5 ± 0.1 & 439.1 &  481.7 & 3.1 & 1.9 \\
$E_{\mathrm{u}}$-4  & 221.7 ± 5.8 & -      & 221.9 ± 5.8 & 9.8* & 9.8* & 384.9  &  387.6 & 3.3 & 3.1 \\
\end{tabular}
\end{ruledtabular}
\end{table*}

Figures~\ref{fig:psidel} and~\ref{fig:MM} depict selected experimental and best match model calculated ellipsometry data for the sample grown on $c$-plane and $m$-plane, respectively. The best match model parameters are summarized in Table~\ref{Table:Ellip}, in comparison to previously reported parameters by Feneberg~\textit{et al.}\cite{Feneberg19aGO} In Figure~\ref{fig:MM}, data from three different rotational azimuth positions of the $m$-sample are included, termed positions P1, P2, and P3. We note that off-block diagonal elements (M$_{13,31}$, M$_{23,32}$, M$_{14}$, and M$_{24}$) are non-zero when the optical axes of the substrate and the thin-film, which are parallel to each other for both samples, are neither parallel nor perpendicular to the plane of incidence. The latter two incidences are observed here approximately in positions P1 and P3, respectively. In position P2, the optical axes of both thin film and substrate, which are perpendicular to  both substrate and film $c$-planes, are oriented approximately 45$^{\circ}$ in between parallel and perpendicular to the plane of incidence. For the $c$-plane sample, the optical axes are always parallel to the plane of incidence, and do not rotate upon sample rotation; hence, the off-block diagonal elements are zero and the standard ellipsometry $\Psi$, $\Delta$ presentation is sufficient in Figure~\ref{fig:psidel}. The ellipsometry data from both samples are analyzed simultaneously by parameterizing $\varepsilon_{\perp,||}$ using the FPSQ model noted above. In this model, two phonon mode pairs were searched for $\varepsilon_{||}$, and four mode pairs were searched for $\varepsilon_{\perp}$, including their respective phonon mode broadening parameters. Additional parameters are the respective high frequency parameters in the FPSQ model, and the film thickness. For the $m$-plane sample we adopted the thickness determined previously by Hilfiker~\textit{et al.} (51.8~nm).\cite{Hilfiker_21aGO} For the $c$-plane sample the thickness was determined in our model analysis ($3.12\pm0.02 \mu$m) which was found in agreement with the growth time estimate ($\approx$3~$\mu$m).

\subsection{Infrared dielectric functions}

\begin{figure*}[t]
\centering
\includegraphics[width=.8\linewidth]{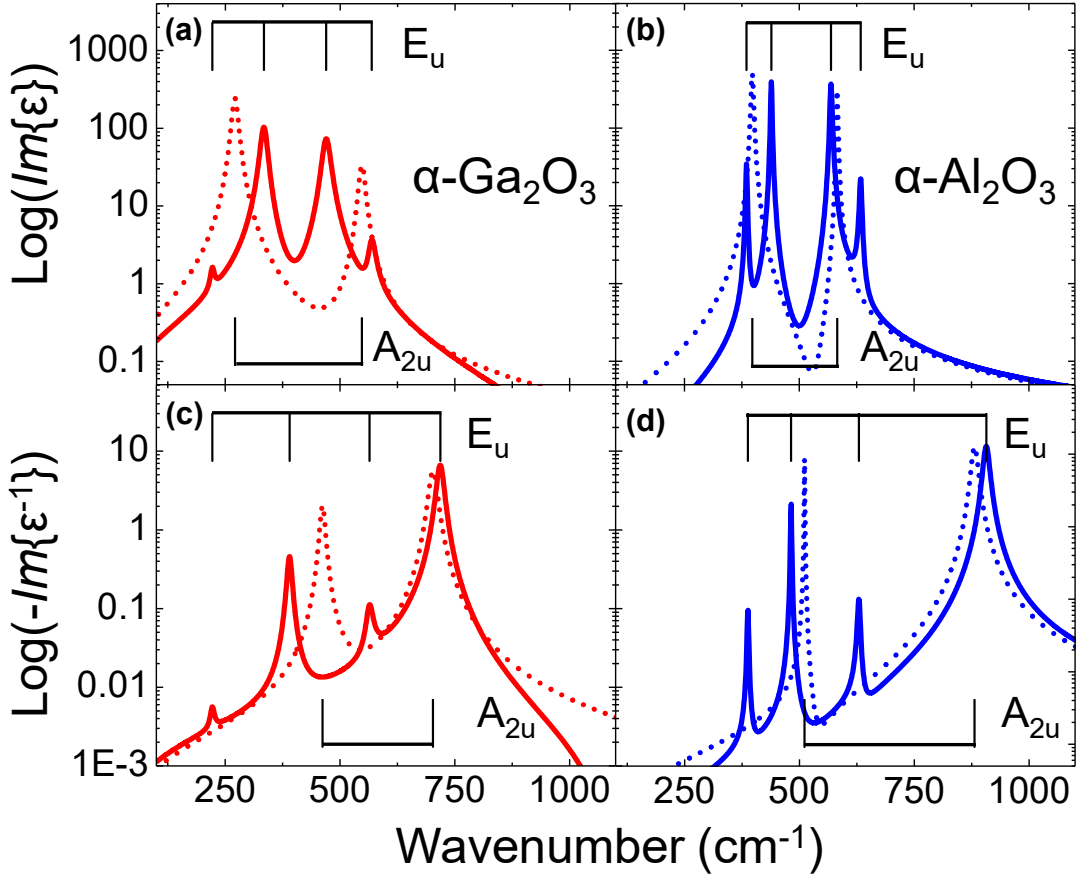}
\caption{\label{fig:dielectriccomp} Imaginary parts of the dielectric functions [panels (a) and (b)] and of the inverse dielectric function [panels (c) and (d)]  for both extraordinary (solid lines) and ordinary (dotted lines) polarization directions, for $\alpha$-Ga$_2$O$_3$ (red lines) and $\alpha$-Al$_2$O$_3$ (blue lines). The latter are replotted from Ref.~\onlinecite{SchubertPRB61_2000}.}
\end{figure*}
The best match model calculated dielectric functions for $\alpha$-Ga$_2$O$_3$ are shown in Figure~\ref{fig:dielectriccomp} in comparison with those for $\alpha$-Al$_2$O$_3$. We show the imaginary parts of the dielectric function and of the negative dielectric loss (inverse dielectric) function. In these presentations, frequency, magnitude, and broadening of TO and LO modes, respectively, are most imminently visible in $\Im\{\varepsilon$\} and $\Im\{-\varepsilon^{-1}$\} for all TO and LO modes, respectively.  All modes are further indicated by the brackets with vertical lines in Figure~\ref{fig:dielectriccomp}. 

\subsection{Static and high frequency dielectric constants}

\begin{table*}\centering 
\caption{\label{Table:epsilon}Static and high frequency dielectric constants of $\alpha$-Ga$_2$O$_3$ from experiment (ellipsometry), and from theory using first principles approaches in different approximations (local density (LDA), generalized gradient (GGA), and linear combination of atomic orbitals (LCAO). Data determined previously from experiment (ellipsometry) for $\alpha$-Al$_2$O$_3$ are included for comparison. Note that high frequency parameters obtained from our ellipsometry analysis are identical with previous results by Hilfiker~\textit{et al.}\cite{Hilfiker_21aGO,Hilfiker_21aAGO} and hence were kept constant here. Note that our static parameters are derived using the IR-active mode parameters and whose error bars propagate then into uncertainty limits for the static parameters.}
\begin{ruledtabular}
\begin{tabular}{{l}{c}{c}{c}{c}{c}{c}{c}{c}{c}}
& \multicolumn{8}{c}{$\alpha$-Ga$_2$O$_3$} & \multicolumn{1}{c}{$\alpha$-Al$_2$O$_3$} \\ \cline{2-9}\cline{10-10}
& \scriptsize{This work} & \scriptsize{This work} & \scriptsize{This work}& \scriptsize{Ref.~\onlinecite{Sharma2021}}&\scriptsize{Ref.~\onlinecite{HePRBaGOtheory2006}} &\scriptsize{Ref.~\onlinecite{Furthmuller_2016}}&\scriptsize{Ref.~\onlinecite{Feneberg19aGO}}&\scriptsize{Ref.~\onlinecite{Hilfiker_21aAGO}} &\scriptsize{Ref.~\onlinecite{SchubertPRB61_2000}}\\
  &Ellip.& LDA & GGA & LDA& LCAO & GGA & Ellip. & Ellip. &Ellip.\\
\hline
$\varepsilon_{\mathrm{DC},||}$ & 18.01$\pm$0.16 &15.54 & 19.12 & 18.67& - & - & - &- & 11.614\\
$\varepsilon_{\mathrm{DC},\perp}$ & 12.14$\pm$0.11& 11.17 & 12.96 & 12.98& - & -& - &- & 9.385\\ 
$\varepsilon_{\infty,||}$ & 3.76 &3.89 & 4.07& 4.46 & 3.03 & 3.8 & 3.65 & 3.76& 3.072\\
$\varepsilon_{\infty,\perp}$ & 3.86 & 3.99 & 4.17 & 4.62 & 3.03& 3.8 & 3.75 & 3.86& 3.077\\ 
		\end{tabular}
\end{ruledtabular}
\end{table*}

Table~\ref{Table:epsilon} lists the high frequency dielectric constants obtained from first principles calculations and experiment performed in this work, in comparison with previous results reported from experiment using IR ellipsometry (Feneberg~\textit{et al.}\cite{Feneberg19aGO}), near-IR to ultra violet spectral range ellipsometry (Hilfiker~\textit{et al.}\cite{Hilfiker_21aGO,Hilfiker_21aAGO}), and from theory at the LDA, linear combination of atomic orbitals (LCAO), and GGA levels by Sharma and Singisetti,\cite{Sharma2021} Orlando~\textit{et al.},\cite{HePRBaGOtheory2006} and Furthm\"{u}ller and Bechstedt.\cite{Furthmuller_2016} Table~\ref{Table:epsilon} further depicts the static dielectric constants for results obtained in this work using the IR-active phonon mode parameters from theory and experiment accordingly, and the LST relationship. Data reported by Sharma and Singisetti are also included. Table~\ref{Table:epsilon} also contains static and high frequency parameters for $\alpha$-Al$_2$O$_3$ reproduced from Ref.~\onlinecite{SchubertPRB61_2000} for further comparison.

\section{Discussion}

Overall, an excellent agreement between experiment and model can be seen for our ellipsometry data. There is increasing noise towards the far-IR range below 200~cm$^{-1}$, and above 1000~cm$^{-1}$ for some of the Mueller matrix spectra. The noise in both regions is caused by loss of reflected beam intensity. For the former, this is caused by the drop in intensity from the globar source (black body radiation), while the latter is caused by reflectance loss due to subtle anisotropy and ambient index of refraction matching for the sapphire substrate. This phenomenon was pointed out and discussed in detail by Schubert, Tiwald, and Herzinger in Ref.~\onlinecite{SchubertPRB61_2000}. Briefly, the recovering dielectric functions for frequencies shortly above the highest IR-active phonon modes in the sapphire substrate cause matching of the sapphire ordinary and extraordinary indices with the ambient index of refraction, and nearly complete loss of reflectance hence. Due to the uniaxial anisotropy this phenomena is polarization dependent, which can be seen in the spectral features occurring at position P2. 

Significant features occur in selected subsets of the experimental data but thereby for all frequencies of the $\alpha$-Ga$_2$O$_3$ IR-active modes, depending on angle of incidence and sample orientation. These features permit us to identify and quantify the model parameters. We note that the lowest mode with $E_{\mathrm{u}}$ symmetry has a very small amplitude, as expected from ours' and others' previous calculations. A subtle feature in our experimental data pertinent to this mode can only be identified in the $c$-plane sample which is highlighted in the inset of Figure~\ref{fig:psidel}. The feature marked by an asterisk is caused by a water absorption line. Hence, the uncertainty limit for TO and LO mode frequencies of mode pair $E_{\mathrm{u}}-4$ is large, and the broadening parameters had to be manually adjusted.   

In general, and as expected for DFT calculations performed using a GGA density functional, our calculated mode frequencies using GGA potentials are slightly lower than the reported experimental frequencies, but otherwise the data sets are in very good agreement. Our GGA results are also typically close to previous results obtained with LDA level calculations. Using LDA functionals, cell volumes are usually lower than recorded by experiment because of an overbinding effect of the local-density approximation, which increases the resulting phonon frequencies. On the other hand, GGA functionals tend to underbind and hence the cell is slightly larger than experimental values. Hence, frequencies from LDA calculations are typically higher than GGA results. In earlier works (Refs.~\onlinecite{Feneberg18aGO,Cusco15,Sharma2021}), however, the gallium semicore 3$d$ states were included into the valence states, which increases the cell volume. In our work, the gallium 3$d$ states are excluded from the valence, and therefore our results obtained with GGA level calculations are close to previous LDA results, while our LDA results clearly stand out.

\paragraph{Phonon mode overlaps.}

The overlap values are of help when comparing the lattice displacements associated with the phonon modes between the two isostructural compounds here. The overlap of a given mode $A$ with another mode $B$, either within the same material or across the two compounds, can be interpreted as the percentage by which the displacement of mode $B$ shares that of mode $A$. This is of interest since it permits us to argue about whether two parental modes, for example, an $A_{\mathrm{1g}}$ mode in $\alpha$-Ga$_{2}$O$_3$ and $\alpha$-Al$_{2}$O$_3$ is anticipated to reveal a one-mode or a more complex mode behavior upon gradual change of composition within an alloy of $\alpha$-(Al$_{x}$Ga$_{1-x}$)$_{2}$O$_3$. We hypothesize that if the overlap between such parental modes is close to unity then one can anticipate that the gradual substitution of Ga by Al does not change the vibrational character of this mode. Hence, this mode should reveal a one mode (single frequency) behavior from one end point to another. We note here again that no experimental data exist that could verify our prediction, and future work is anticipated to provide further clarification.   

\paragraph{Raman-active and silent modes.}

The eigenvector overlaps for Raman-active $A_{\mathrm{1g}}$, $E_{\mathrm{g}}$, and $A_{\mathrm{2g}}$, respectively, are shown in Tables~\ref{tab:ag_overlaps},~\ref{tab:eg_overlaps}, and~\ref{tab:a2g_overlaps}. We observe that both $A_{\mathrm{1g}}$ modes fully maintain their displacement character, i.e., the dot products between the eigenvectors of both modes with the same mode index between the two compounds is unity. A very small portion is shared crosswise, i.e., the second (first) mode in sapphire contains about a few percent of the displacement features of the first (second) mode in gallium oxide. Note that the vector coordinates, in general, in the two symmetry sub-spaces for the two binary compounds are not exactly the same. Hence, we predict that the $A_{\mathrm{1g}}$ modes will display a single mode behavior and shift linearly throughout the composition range. The same is true for the silent modes (Table~\ref{tab:a1u_overlaps}). The results for $E_{\mathrm{g}}$ are much more diverse. Mode $E_{\mathrm{g}}-1$ should clearly depict a single-mode behavior. However, the remaining 4 modes change their character substantially. We highlight the pair of modes $E_{\mathrm{g}}-3$ and $E_{\mathrm{g}}-5$, where the latter in sapphire has attained almost all of the character of mode $E_{\mathrm{g}}-3$ in gallium oxide. We predict that these modes may show a complex behavior upon alloying, especially since the character of mode $E_{\mathrm{g}}-4$ must somehow cross over with mode $E_{\mathrm{g}}-3$. We note, for example, the normalization of overlaps for mode $E_{\mathrm{g}}$-1 in $\alpha$-Ga$_{2}$O$_3$ with all modes $E_{\mathrm{g}}$-1,2,3,4,5 in $\alpha$-Al$_{2}$O$_3$, $0.99830^2 + 0.01711^2 + 0.03720^2 + 0.01792^2 + 0.03753^2=1.00001$ in Table~\ref{tab:eg_overlaps}, also indicating the numerical uncertainty limits in our computation analysis. 

\paragraph{Infrared-active modes.}

The overlaps for TO and LO modes with $A_{\mathrm{2u}}$ and $E_{\mathrm{u}}$ symmetries, respectively, are shown in Tables~\ref{tab:a2u_overlaps} and~\ref{tab:eu_overlaps}. Here we included also overlaps between TO and LO modes within and across the same mode index as well as between the two binary compounds.  Let us start with a relatively simple case of $A_{\mathrm{2u}}$ modes. The overlap matrices for TO and LO modes, respectively, between aAO and aGO are given in Table~\ref{tab:a2u_overlaps}. What these overlaps tell us is that the two $A_{\mathrm{2u}}$ modes have respectively very similar eigenvectors (atomic displacement patterns) between the two materials. The overlaps of respective TO modes between $\alpha$-Ga$_{2}$O$_3$ and $\alpha$-Al$_{2}$O$_3$,
and similarly so for the overlaps of LO modes, have overlap values $>0.95$. At the same time, the mixed overlaps, e.g., the overlap of mode 1 for gallium oxide and mode 2 for sapphire, and~\emph{vice versa}, have low values ($<0.26$) indicating that these two $A_{\mathrm{2u}}$ modes remain very distinct from one another upon alloying. If these modes had higher overlap values, it could be reasonable to predict a more complex mode evolution than one-mode behavior with alloying. Note that for the same material the eigenvectors are orthogonal and the overlaps are unity. In contrast, when we look at the overlap matrices between TO and LO modes for each of the materials separately, i.e., within the same compound, we observe that each mode has a mixed character, here with an equal level of similarity of $\approx 0.7$ to both TO modes (or 0.67 when one compares the sapphire LO modes to the gallium oxide TO modes). The $A_{\mathrm{2u}}$ TO modes maintain their character (0.98 and 0.96 overlaps) when transitioning between $\alpha$-Ga$_{2}$O$_3$ and $\alpha$-Al$_{2}$O$_3$, and the same can be said for the LO modes (0.95 and 0.96 overlaps). However, there is also a contribution from mode character crossing of 0.18 for the TO and 0.25 for the LO modes. Hence, a clear prediction for the mode character upon gradual alloying may be difficult to make and one could expect to see a one mode behavior for all four modes.

The $E_{\mathrm{u}}$ modes exhibit a more complex picture. Overall, TO and LO modes mostly maintain their character, e.g., the diagonal (equal mode index) overlaps for both alloys between TO modes are 0.99, 0.97, 091, and 0.92 and between LO modes are 0.97, 1.00, 0.96, and 0.96. Thus, a linear shift with single mode behavior may be seen for all modes. However, there is also substantial admixture seen of mode character for mode 4 in sapphire which also overlaps approximately 36\% with mode 3 and 15\%  with mode 2 in gallium oxide. That being said, mode 4 in sapphire is still 92\% similar to mode 4 in gallium oxide. Likewise, TO mode 3 in sapphire resembles 81\% of LO mode 3 in gallium oxide, and it could be anticipated that a defect-like mode pair develops between the two modes of the two compounds. While these modes each maintain high similarity (96\% each) with their complementary modes, it could very well be alloying disrupts the crystal structure enough to present as a defect-like mode rather than follow a one-mode evolution.  We note further that both mode pairs TO-1 and LO-2 in both compounds resemble each other closely, within and across the compounds, which can be well understood by the small frequency differences formed by each pair (DFT-GGA: aGO: TO-1 = 546.64~cm$^{-1}$, LO-2 = 544.68~cm$^{-1}$; aAO: TO-1 = 601.58~cm$^{-1}$, LO-2 =  598.17~cm$^{-1}$. GSE: aGO: TO-1 = 568.5~cm$^{-1}$, LO-2 =  564.3~cm$^{-1}$; aAO: TO-1 =  633.6~cm$^{-1}$, LO-2 =  629.5~cm$^{-1}$). We note finally, for example, the normalization of overlaps for mode $E_{\mathrm{u}}$-4 (LO) in $\alpha$-Ga$_{2}$O$_3$ with all modes $E_{\mathrm{u}}$-1,2,3,4 (LO) in $\alpha$-Al$_{2}$O$_3$, $0.01261^2 + 0.02565^2 + 0.28113^2 + 0.95924^2=0.999992$ in Table~\ref{tab:eu_overlaps}, and likewise with all modes $E_{\mathrm{u}}$-1,2,3,4 (TO) in $\alpha$-Al$_{2}$O$_3$, $0.03766^2 + 0.10302^2 + 0.37239^2 + 0.92157^2=0.999997$, also indicating the numerical uncertainty limits in our computation analysis. 

\paragraph{Raman mode shifts.}

Between the GGA level results, we can formulate the anticipated Raman frequency shifts ($\omega_{\mathrm{RA}}$[$\alpha$-Al$_{2}$O$_3$]-$\omega_{\mathrm{RA}}$[$\alpha$-Ga$_{2}$O$_3$] ) between the two materials as a function of composition ($\Delta A_{\mathrm{1g}}-1 =59.10$~cm$^{-1}$, $\Delta A_{\mathrm{1g}}-2 =174.03$~cm$^{-1}$, $\Delta E_{\mathrm{g}}-1 =5.3 $~cm$^{-1}$, $\Delta E_{\mathrm{g}}-2 =115.94 $~cm$^{-1}$, $\Delta E_{\mathrm{g}}-3 = 86.95 $~cm$^{-1}$, $\Delta E_{\mathrm{g}}-4 =102.47 $~cm$^{-1}$, $\Delta E_{\mathrm{g}}-5 =121.96 $~cm$^{-1}$). The shifts are all positive, with those for modes $A_{\mathrm{1g}}-2$ and $E_{\mathrm{g}}-1$ largest and smallest, respectively. 

\paragraph{Infrared dielectric functions.}

Overall, the phonon mode picture for the IR-active modes between the two isostructural compounds, at first sight (Figure~\ref{fig:dielectriccomp}), is very similar. The TO modes for the ordinary direction show two strong modes flanked by two rather weak modes, one at the far-IR side and one at the mid-IR side (red and blue solid lines in $\Im\{\varepsilon_{\perp}\}$). The extraordinary direction shows two rather strong modes (dotted lines) and both line shapes shift towards the longer wavelengths upon substitution by the heavier element Ga, most evidently seen in Figure~\ref{fig:dielectriccomp}. A similar behavior is seen for the LO modes, except that second and fourth mode for the ordinary direction have considerably smaller amplitudes. Furthermore, a noticeable increase in broadening is associated with the shift. We believe that this increase is due to lesser crystal quality of the thin films compared with the high structural quality bulk sapphire substrates. However, the crystal quality of the thin-film sample is still very high, as detailed in Ref.~\onlinecite{jinno2020crystal}. Note that the mode broadening parameters are, as a good estimate, directly proportional to the width of the peaks seen in Figure~\ref{fig:dielectriccomp}. The red-shift is not associated with a substantial spectral spread or compression between the individual modes which indicates that the TO-LO splitting between the associated mode TO and LO mode pairs in $\alpha$-Ga$_2$O$_3$ is very similar to $\alpha$-Al$_2$O$_3$. In other words, the overall polarity, i.e., the ability of the lattice to respond to IR radiation, is very similar for both compounds. 

\paragraph{Infrared-active mode shifts.}

Between the GGA level results, we can formulate the anticipated frequency shifts ($\omega_{\mathrm{TO,LO}}$[$\alpha$-Al$_{2}$O$_3$]-$\omega_{\mathrm{TO,LO}}$[$\alpha$-Ga$_{2}$O$_3$] ) between the two materials as a function of composition for TO modes ($\Delta_{\mathrm{TO}} A_{\mathrm{2u}}-1 =32.13$~cm$^{-1}$, $\Delta_{\mathrm{TO}} A_{\mathrm{2u}}-2 =108.17$~cm$^{-1}$, $\Delta_{\mathrm{TO}} E_{\mathrm{u}}-1 =54.93 $~cm$^{-1}$, $\Delta_{\mathrm{TO}} E_{\mathrm{u}}-2 = 85.67 $~cm$^{-1}$, $\Delta_{\mathrm{TO}} E_{\mathrm{u}}-3 =89.15 $~cm$^{-1}$, $\Delta_{\mathrm{TO}} E_{\mathrm{u}}-4 = 144.92 $~cm$^{-1}$) and for LO modes ($\Delta_{\mathrm{LO}} A_{\mathrm{2u}}-1 =163.74$~cm$^{-1}$, $\Delta_{\mathrm{LO}} A_{\mathrm{2u}}-2 =46.86$~cm$^{-1}$, $\Delta_{\mathrm{LO}} E_{\mathrm{u}}-1 = 179.28$~cm$^{-1}$, $\Delta_{\mathrm{LO}} E_{\mathrm{u}}-2 = 51.98$~cm$^{-1}$, $\Delta_{\mathrm{LO}} E_{\mathrm{u}}-3 = 77.98$~cm$^{-1}$, $\Delta_{\mathrm{LO}} E_{\mathrm{u}}-4 = 140.91$~cm$^{-1}$). The shifts are all positive, with those for TO modes $E_{\mathrm{u}}-4$ and $A_{\mathrm{2u}}-1$ largest and smallest, respectively, and LO modes $E_{\mathrm{u}}-1$ and $E_{\mathrm{u}}-2$ largest and smallest, respectively.

Table~\ref{Table:Ellip} compares existing experimental results for $\omega_{\mathrm{TO}}$ reported by Feneberg \textit{et al.} with our results. Largely, these results are consistent with our best-match model parameters. We note further that the mode at the far-IR end with $A_{\mathrm{2u}}$, not previously observed, has a very small amplitude and is associated with a very small TO-LO splitting. No LO modes have been reported prior to this work. We can thus evaluate the shift of the IR-active mode frequencies between the two alloys for TO modes, i.e., $A_{\mathrm{2u}}-1 =35.31(32.13)$~cm$^{-1}$, $A_{\mathrm{2u}}-2 =126.7(108.17)$~cm$^{-1}$, $E_{\mathrm{u}}-1 =65.1(54.93)$~cm$^{-1}$, $E_{\mathrm{u}}-2 = 99.5(85.67)$~cm$^{-1}$, $E_{\mathrm{u}}-3 =105.1(89.15)$~cm$^{-1}$, $E_{\mathrm{u}}-4 = 163.2(144.92)$~cm$^{-1}$, and for LO modes, i.e., $A_{\mathrm{2u}}-1 =178.8(163.74)$~cm$^{-1}$, $A_{\mathrm{2u}}-2 =49.1(46.86)$~cm$^{-1}$, $E_{\mathrm{u}}-1 =188.3(179.28)$~cm$^{-1}$, $E_{\mathrm{u}}-2 = 65.2(51.98)$~cm$^{-1}$, $E_{\mathrm{u}}-3 = 91.7(77.98)$~cm$^{-1}$, $E_{\mathrm{u}}-4 = 165.7(140.91)$~cm$^{-1}$, where the values in brackets are those from our DFT results. A remarkable agreement between difference results measured by ellipsometry and calculated by DFT is observed here. Likewise, the experimentally obtained TO modes $E_{\mathrm{u}}-4$ and $A_{\mathrm{2u}}-1$ shift largest and smallest, respectively, and LO modes $E_{\mathrm{u}}-1$ and $E_{\mathrm{u}}-2$ shift largest and smallest, respectively, again in excellent agreement with the predicted shifts from DFT results.

\paragraph{Static and high-frequency dielectric constants.}

The high-frequency dielectric constants for $\alpha$-Ga$_{2}$O$_3$ $\varepsilon_{\infty,\parallel}$ and $\varepsilon_{\infty,\perp}$ were found identical with the previously reported values by Hilfiker \textit{et al.}, $\varepsilon_{\infty,\parallel} = 3.76$ and $\varepsilon_{\infty,\perp} = 3.86$. Using the LST relation and frequencies in our complete set of TO and LO modes, we find the static dielectric constants $\varepsilon_{\mathrm{DC},\parallel} = 18.01 \pm 0.16$ and $\varepsilon_{\mathrm{DC},\perp} = 12.14\pm 0.11 $. Experimental values for the static dielectric constants can not be found in the existing literature. Theoretical work by Sharma and Singisetti report the values found via the LST relation as $\varepsilon_{\mathrm{DC},\parallel} = 18.67$  and $\varepsilon_{\mathrm{DC},\perp} = 12.98 $,\cite{Sharma2021} which are in very close agreement with our experimental results as well as GGA level calculations (19.12 and 12.96, respectively). Overall, both static and high frequency limits decrease substantially from $\alpha$-Ga$_{2}$O$_3$ to $\alpha$-Al$_{2}$O$_3$, which leads to a reduction in index of refraction, as revealed recently by Hilfiker \textit{et al.} from analysis of epitaxial thin films.\cite{Hilfiker_21aAGO} The reduction can be associated with the smaller atomic size of aluminum versus gallium.

\section{Conclusion}

A combined experimental and theoretical analysis of IR dielectric and phonon mode properties of corundum structure gallium oxide revealed similarities and differences introduced upon the replacement of aluminum with gallium. Firstly, we identified the complete set of IR-active phonon modes in gallium oxide. These results compared reasonably well to existing partial data and to results from calculations. Secondly, all gallium oxide Raman-active and silent modes from theory are compared with $\alpha$-Al$_{2}$O$_3$  and with results from experiment. We find that changes in mode frequencies observed by experiments are highly consistent with our theoretical calculations. Thirdly, mode eigenvector overlaps provide insight into similarities and differences of phonon modes in both binary compounds. We suggest use of this information to identify the mode character upon alloying in the ternary system of corundum structure aluminum gallium oxide.    

\acknowledgments

This work was supported in part by the National Science Foundation (NSF) under awards NSF DMR 1808715 and NSF/EPSCoR RII Track-1: Emergent Quantum Materials and Technologies (EQUATE), Award OIA-2044049, and by Air Force Office of Scientific Research under awards FA9550-18-1-0360, FA9550-19-S-0003, and FA9550-21-1-0259, and by ACCESS, an AFOSR Center of Excellence, under award FA9550-18-1-0529, and by the Knut and Alice Wallenbergs Foundation award 'Wide-bandgap semiconductors for next generation quantum components'. M.~S. acknowledges the University of Nebraska Foundation and the J.~A.~Woollam~Foundation for support. R.~J. acknowledges the supported by JSPS Overseas Challenge Program for Young Researchers 1080033. This work was also supported in part by the Swedish Research Council VR award No. 2016-00889, the Swedish Foundation for Strategic Research Grant Nos. RIF14-055 and EM16-0024, by the Swedish Governmental Agency for Innovation Systems VINNOVA under the Competence Center Program Grant No. 2016–05190, and by the Swedish Government Strategic Research Area in Materials Science on Functional Materials at Link{\"o}ping University, Faculty Grant SFO Mat LiU No. 2009-00971. 

The data that support the findings of this study are available from the corresponding author upon reasonable request.
\bibliography{aGOIRphonon}
\end{document}